\documentclass[smallextended]{svjour3}       
\usepackage[english]{babel}
\usepackage{amssymb}
\usepackage{amsbsy}
\usepackage{amsmath}
\usepackage{color}
\newcommand{\vx}{{\mathbf x}}
\newcommand{\vv}{{\mathbf v}}
\newcommand{\va}{{\mathbf a}}
\newcommand{\vu}{{\mathbf u}}

\smartqed  

\usepackage{mathtools}

\usepackage{soul}
\usepackage[normalem]{ulem}
\usepackage{verbatim}
\usepackage{subfigure}

\newcommand{\be}{\begin{equation}}
\newcommand{\ee}{\end{equation}}
\newcommand{\ba}{ {\bf a}}
\newcommand{\bx}{ {\bf x}}
\newcommand{\by}{ {\bf y}}

\DeclareMathOperator{\sgn}{sgn}

\usepackage{lineno}
\journalname{Computational Geosciences}

\begin{document}

\title{Machine learning for graph-based representations of three-dimensional discrete fracture networks}

\titlerunning{Machine learning for 3D DFN}
\authorrunning{Valera, et al.}



\author{ Manuel~Valera \and Zhengyang~Guo \and Priscilla~Kelly \and Sean~Matz \and Vito~Adrian~Cantu \and Allon~G.~Percus\and Jeffrey~D.~Hyman \and Gowri~Srinivasan \and Hari~S.~Viswanathan}

\institute{
M.~Valera \at
Computational Science Research Center, San Diego State University, San Diego, CA 92182, United States of America, and Institute of Mathematical Sciences, Claremont Graduate University, Claremont, CA 91711, United States of America,
\and
Z. Guo \at
Data \& Decision Sciences, C2FO, Leawood, KS 66206, United States of
America
\and
P. Kelly \at
Computational Science Research Center, San Diego State University, San Diego, CA 92182, United States of America, and Institute of Mathematical Sciences, Claremont Graduate University, Claremont, CA 91711, United States of America,
\and
S. Matz \at
Institute of Mathematical Sciences, Claremont Graduate University, Claremont, CA 91711, United States of America
\and 
V. A. Cantu \at
Computational Science Research Center, San Diego State University, San Diego, CA 92182, United States of America, and Institute of Mathematical Sciences, Claremont Graduate University, Claremont, CA 91711, United States of America,
\and 
A. G. Percus \at 
Institute of Mathematical Sciences, Claremont Graduate University, Claremont, CA 91711, United States of America
\and 
J. D. Hyman \at
Computational Earth Science Group (EES-16), Earth and Environmental Sciences Division, Los Alamos National Laboratory, Los Alamos, NM 87545, United States of America
\emph{Corresponding Author} \email{jhyman@lanl.gov}  
\and 
G. Srinivasan \at
Applied Mathematics and Plasma Physics (T-5), Theoretical Division, Los Alamos National Laboratory, Los Alamos, NM 87545, United States of America
\and 
H. S. Viswanathan
\at
Computational Earth Science Group (EES-16), Earth and Environmental Sciences Division, Los Alamos National Laboratory, Los Alamos, NM 87545, United States of America}

\maketitle

\begin{abstract}
Structural and topological information play a key role in modeling flow and transport through fractured rock in the sub-surface.  
Discrete fracture network (DFN) computational suites such as {\sc dfnWorks}~\cite{Hyman2015} are designed to simulate flow and transport in such porous media.  Flow and transport calculations reveal that a small backbone of fractures exists, where most flow and transport occurs.  Restricting the flowing fracture network to this backbone provides a significant reduction in the network's effective size. However, the particle tracking simulations needed to determine the reduction are computationally intensive. Such methods may be impractical for large systems or for robust uncertainty quantification of fracture networks, where thousands of forward simulations are needed to bound system behavior.

In this paper, we develop an alternative network reduction
approach to characterizing transport in DFNs, by combining graph theoretical and machine learning methods. We consider a graph representation where nodes signify fractures and edges denote their intersections. Using random forest and support vector machines, we rapidly identify a subnetwork that captures the flow patterns of the full DFN, based primarily on node centrality features in the graph.  Our supervised learning techniques
train on particle-tracking backbone paths found by {\sc dfnWorks}, but run in negligible time compared to those simulations.  We find that our predictions
can reduce the network to approximately 20\% of its original size, while still generating breakthrough curves consistent with those of the original network.

\end{abstract}

\keywords{Machine learning \and Discrete Fracture Networks \and Support Vector Machines \and Random Forest \and Centrality  }


\section{Introduction}
\label{S:intro}

In low permeability media, such as shales and granite,  interconnected networks of fractures are the primary pathways for fluid flow and associated transport of dissolved chemicals.  Characterizing flow and transport through fractured media in the subsurface is critical in many civil, industrial, and security applications including drinking water aquifer management~\cite{national1996rock}, hydrocarbon extraction~\cite{hyman2016understanding,Karra2014}, and carbon sequestration~\cite{jenkins2015state}.
For example, increasing extraction efficiency from hydraulic fracturing, preventing leakage from CO$_2$ sequestration, or detecting the arrival time of subsurface gases from a nuclear test requires a model that can accurately simulate flow and transport through a subsurface fracture network.  

In these sparse systems fracture network topology controls system behavior but is uncertain because the exact location of subsurface fractures cannot be determined with sites often 1000s of feet below the ground. This necessitates a method to quickly calculate the transport time of solutes through realistic statistical representations of  fracture networks.
The topology of the network can induce flow channeling,  where isolated regions of high velocity form within the network~\cite{abelin1991large,abelin1985final,dreuzy2012influence,frampton2011numerical,hyman2015influence,rasmuson1986radionuclide}.
The formation of these flow channels indicates that much of the flow and transport occurs in a subnetwork of the whole domain. 
There are techniques available to identify the fractures that make up these subnetworks, commonly referred to as the backbone~\cite{aldrich2016analysis,maillot2016connectivity}.
However, these techniques require resolving flow and/or transport through the entire network prior to being able to identity the backbone. 
For large networks, particle-based simulations~\cite{robinson2010particle,srinivasan2010random} can be costly in terms of required computational time. 
These costs are exacerbated because numerous network realizations are required to obtain trustworthy statistics of upscaled quantities of interest, e.g., the distribution of fracture characteristics that make up the backbone. 
Because the connectivity of the networks is dominant in determining where flow and transport occurs in sparse systems than geometric or hydraulic properties~\cite{hyman2016fracture}, it should be possible to identify high-flow and transport subnetworks
using the network's topological properties.

Graph representations of fracture networks have been proposed by Ghaffari, et al.~\cite{ghaffari2011} and independently by Andresen, et al.~\cite{Andresen2013}.  These graph mappings allow for a characterization of the network topology of both two- and three-dimensional fracture systems, and moreover enable quantitative comparisons between real fracture networks and models generating synthetic networks.  Vevatne, et al.~\cite{Vevatne2014} and Hope, et al.~\cite{Hope2015} have used this graph construction for analyzing fracture growth and propagation, showing how topological properties of the network such as assortativity relate to the growth mechanism.
Hyman, et al.~\cite{hyman2017accurate} used graph representations of three-dimensional fracture networks to isolate subnetworks where the fastest transport occurred by finding the shortest path between inflow and outflow boundaries.  
Santiago, et al.~\cite{Santiago2013,Santiago2014,Santiago2016}  proposed a method of topological analysis using a related graph representation of fracture networks.  By measuring centrality properties of nodes in the graph, which describe characteristics such as the number of shortest paths through a given node, they developed a method intended to predict regions of high flow conductivity in the network.

In recent years, there has been increased interest in the use of machine learning in the geosciences.  A range of different regression and classification methods have been applied to a model of landslide susceptibility, demonstrating their predictive value~\cite{goetz_2015}.  Community detection methods have been used in fractured rock samples to identify regions expected to have high flow conductivity~\cite{Santiago2014}.  Clustering analysis has been used in subsurface systems to construct more accurate flow inversion algorithms~\cite{mud_2016}.

We combine the two approaches of discrete fracture network (DFN) graph representations and machine learning to identify subnetworks that conduct significant flow and transport.
We represent a fracture by a node in the graph, and an intersection between two fractures by an edge.
Using this construction, the graph retains topological information about the network as node-based properties, or ``features.''
On the basis of six features, four topological and two physical, we apply machine learning to reduce the fracture network to a subnetwork.
We use two supervised learning methods, random forest and support vector machines, that train on backbones defined using particle-tracking simulations in the entire DFN.
Both algorithms have the advantage of being general-purpose methods, suitable for geometric as well as non-geometric features, and requiring relatively little parameter tuning. 
\textit{Our overall goal is to combine graph theoretical methods and machine learning to isolate subnetworks where the the majority of flow and transport occurs, as an alternative to high-fidelity discrete fracture network flow and transport simulations on the entire fracture network.}

Although our algorithms train on particle backbones, we depart from a conventional machine learning approach in that we do not necessarily aim to reproduce these backbones.  The objective is to learn from them, obtaining network reductions that are valid for characterizing flow and transport.  The particle backbone is only one out of many possible such reductions.  Ultimately, the quality of a result is measured through its breakthrough curve (BTC), which gives the distribution of times for passive tracer particles to pass through the network.

Under different parameter choices for random forest and SVM, we are able to reduce fracture networks on average to between 39\% and 2.5\% of their original number of fractures.
Reductions to as little as 21\% 
still result in a BTC in good agreement with that of the full network.  Thus, our methods yield subnetworks that are significantly smaller than the full network, while matching its main flow and transport properties.  Notably, we are able to generate these subnetworks in seconds, whereas the computation time for extracting the backbone from particle-based transport simulations is on the order of an hour.  

We also assess the importance of the different features used to characterize the data, finding that they cluster into three natural groups.  The global topological quantities are the most significant ones, followed by the one local topological quantity we use. The physical quantities are the least significant ones, though still necessary for the performance of the classifier.

In section~\ref{sec:DFN}, we describe the flow and transport simulations used to determine particle-trace based backbones in the DFN.  Section~\ref{S:grdfn} describes the graph representation, as well as the features used to characterize nodes in the graph.  Section~\ref{S:methods} discusses the details of the machine learning methods used, and section~\ref{S:results} presents the results of these methods.  Finally, in section~\ref{S:conc}, we discuss the implications of our results and provide conclusions.

\section{Discrete Fracture Network}\label{sec:DFN}

Discrete fracture networks (DFN) models are one common simulation tools used to investigate flow and transport in fractured systems. 
In the DFN methodology, the fracture network and hydrological properties are explicitly represented as discrete entities within an interconnected network of fractures. 
The inclusion of such detailed structural and hydrological properties allows DFN models to represent a wider range of transport phenomena than traditional  continuum models~\cite{painter2005upscaling,painter2002power}.
In particular,  topological, geometric, and hydrological characteristics can be directly linked to physical flow observables.

We use the computational suite {\sc dfnWorks}~\cite{Hyman2015} to generate each DFN, solve the steady-state flow equations and determine transport properties.
 {\sc dfnWorks} combines the feature rejection algorithm for meshing ({\sc fram})~\cite{hyman2014conforming}, the LaGriT meshing toolbox~\cite{lagrit2011}, the parallelized subsurface flow and reactive transport code {\sc pflotran}~\cite{lichtner2015pflotran}, and an extension of the {\sc walkabout} particle tracking method~\cite{makedonska2015particle,painter2012pathline}.  
{\sc fram} is used to generate three-dimensional fracture networks.  
LaGriT is used to create a computational mesh representation of the DFN in parallel.
 {\sc pflotran} is used to numerically integrate the governing flow equations.
  {\sc walkabout} is used to determine pathlines through the DFN and simulate solute transport.
 Details of the suite, its abilities,  applications,  and references for detailed implementation are provided in~\cite{Hyman2015}.

\begin{figure}[tb!] \centerline{
 \begin{tabular}{ccc}
 (a)  &  &  (b)   \\
 \includegraphics[width=0.45\textwidth]{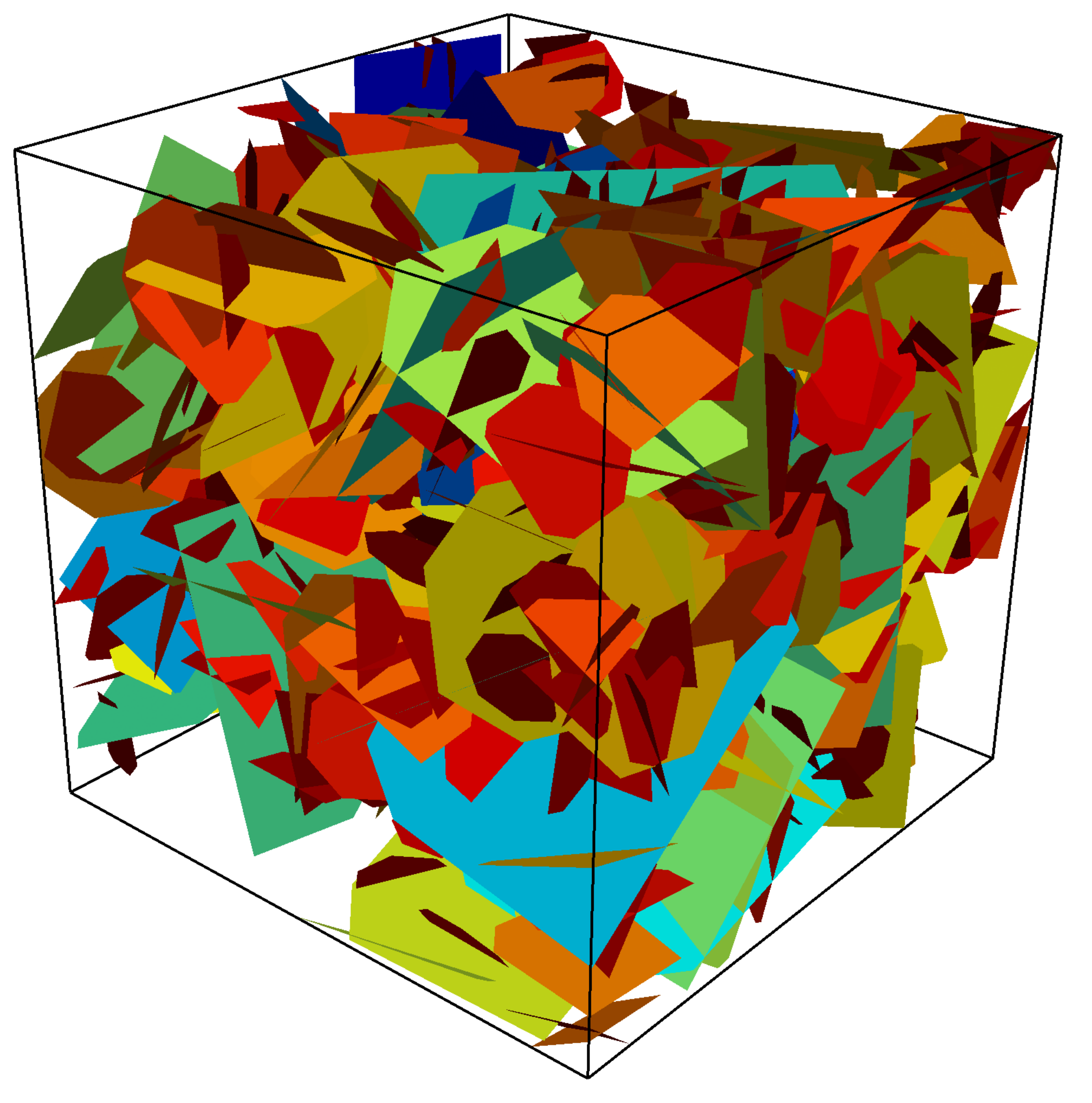} &\qquad & \includegraphics[width=0.45\textwidth]{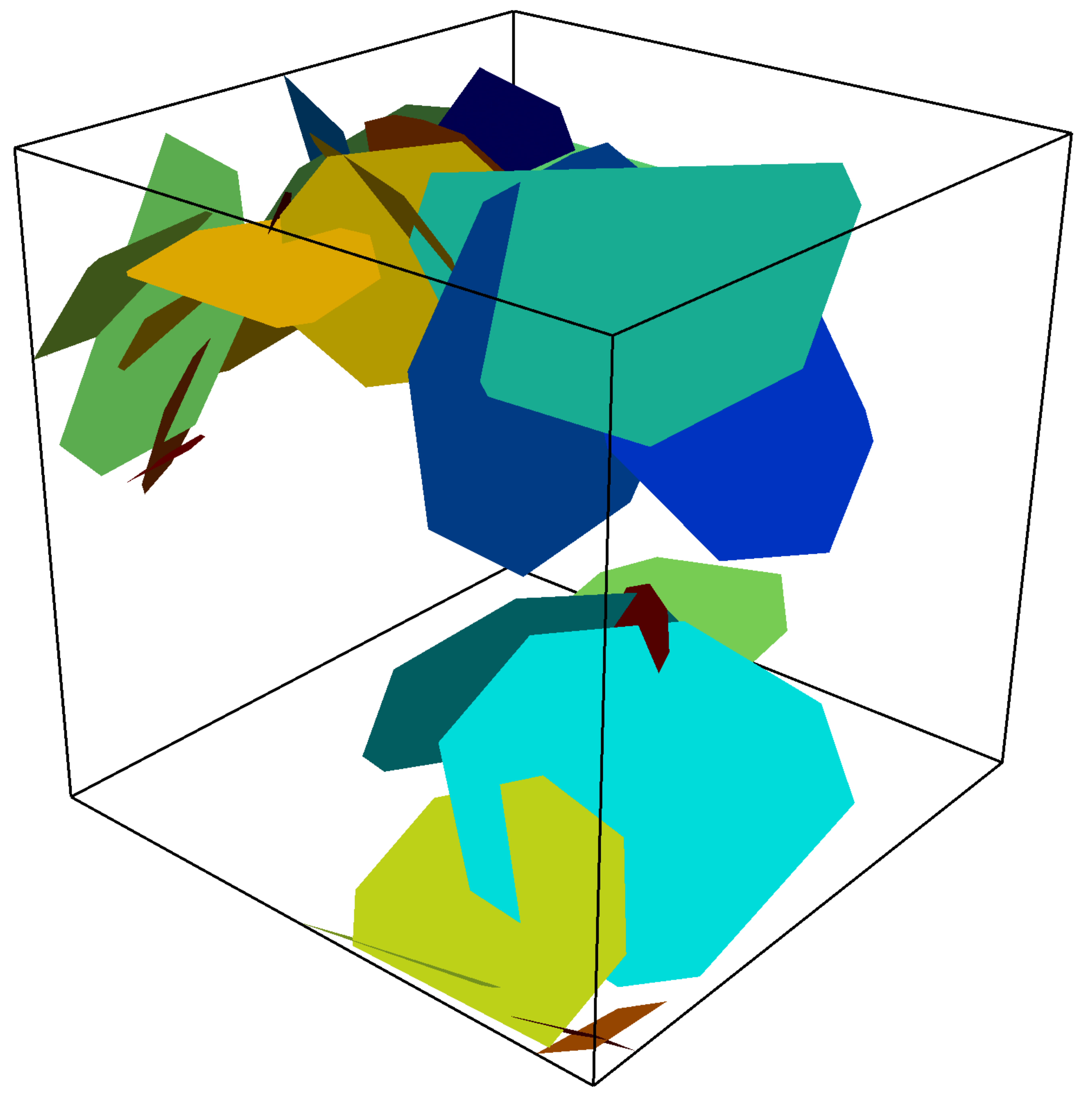} \\
 \end{tabular}}
\caption{\label{fig:dfn} (a) A DFN composed of 499 fractures. (b) Backbone extracted from (a) using particles trajectories where a majority of mass transport occurs.  Inlet plane is shown on front left; outlet plane is on rear right.}
\end{figure}

One hundred generic networks, composed of circular fractures with uniformly random orientations, are generated. 
Each DFN lies within a cubic domain with sides of length 15 meters.
The fracture radii $r$ [m] are sampled from a truncated power law distribution with exponent $\alpha$ and upper and lower  cutoffs ($r_u$; $r_0$).
We select a value of $\alpha = 2.6$ so that the distribution has finite mean and variance.
The lower cut off $r_0$ is set to one meter and the upper cut off equal $r_u$ is set to five meters. 
Fracture centers are sampled uniformly throughout the domain. 
The networks are fairly sparse, with an average P$_{32}$ value (fracture surface area over total volume) of 1.97  [m$^{-1}$] and variance 0.03. 
In all networks, at least one set of fractures connects the inflow and outflow boundaries. 
This constraint removes isolated clusters that do not contribute to flow. 
An example of one fracture network is shown in Figure~\ref{fig:dfn}(a).
On average, meshing each full-network, solving for flow, and tracking particles takes around 30 minutes of wall clock time.
Timing for these computations was performed using a server that has 64 cores; 1.4 GHz  AMD Opteron(TM) Processor 6272 with 2048 KB of cache each.
Meshing and flow simulations are performed in parallel using 16 cores. 
Transport is performed using a single core.

Within each network, the governing equations are numerically integrated to obtain a steady-state pressure field. 
Purely advective passive particles are tracked through steady-state flow fields to simulate transport and their pathlines correspond to the fluid velocity field. 
Details of the flow and transport simulations are provided in the appendix. 
Particle pathlines are used to identify backbones in the DFN, connected subsets of fractures where a substantial portion of flow and transport occurs, using the methods of Aldrich, et al.~\cite{aldrich2016analysis}.
In their method, membership in the backbone is the result of a large amount of mass passing through particular pathways of connected fractures. Using this definition of backbone, the breakthrough curve, e.g., travel time distributions, on subnetwork defined by the backbone is not guaranteed to match the breakthrough curve on the full network, but it does identify primary flow and transport paths in the system.
Figure~\ref{fig:dfn}(b) shows the backbone extracted from the network shown in Figure~\ref{fig:dfn}(a).

\section{Graph Representation}
\label{S:grdfn}

\subsection{Graph Formation}

We construct a graph representation of each DFN based on the network topology using the method described in Hyman, et al.~\cite{hyman2017accurate}.
For every fracture in the DFN, there is a unique node in a graph.  
If two fractures intersect, then there is an edge in the graph connecting the corresponding nodes.
This mapping between DFN and graph naturally assigns fracture-based properties, both geometric and hydrological, as node attributes.
Edges are assigned unit weight to isolate topological attributes from other attributes that could be considered.

\begin{figure}[tb!]
\vspace{-12pt}
\centering\includegraphics[width=0.95\linewidth]{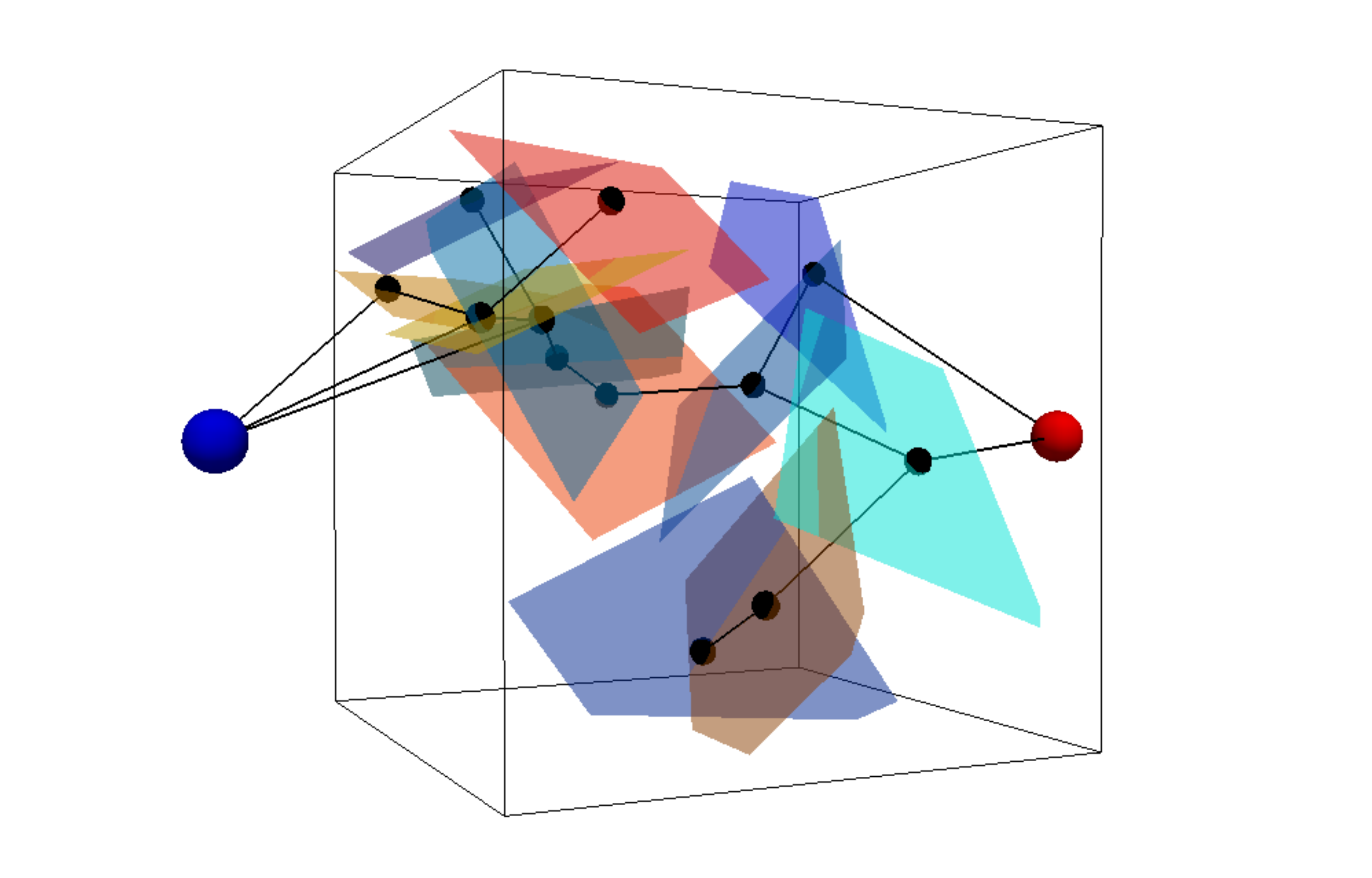}
\vspace{-18pt}
\caption{Graph representation of a twelve-fracture discrete fracture network in three dimensions.  Fractures (semi-transparent colored planes) are represented as nodes (black), intersections between fractures are represented by edges in the graph (solid black lines). A source node (blue) and a target node (red) are also included into the graph to include boundary conditions and flow direction. }\label{fig:dfn2graph3D}
\end{figure}

Source and target nodes are included into the graph to incorporate flow direction.
Every fracture that intersects the inlet plane is connected to the source node and every fracture that intersects to the outlet plane is connected to the target node.
The inclusion of flow direction is essential to identify possible transport locations, which depend upon the imposed pressure gradient~\cite{neuman2005trends}.
An example of this mapping for a three-dimensional twelve fracture network is shown in  figure \ref{fig:dfn2graph3D}.  
Each of the fractures (semi-transparent colored planes) are represented as nodes (black), intersections between fractures are represented by edges in the graph (solid black lines).
The source node is colored blue and the target node is colored red. 

Every subgraph has a unique pre-image in the fracture network that is a subnetwork of the full network because the mapping between the network and graph is bijective.
Thus, flow and transport simulations can be performed on these subnetworks, and compared to results obtained on the full networks.

\subsection{Node features}
\label{S:features}

\begin{figure}
\centering
\begin{minipage}{0.48\linewidth}
\centering
\includegraphics[width=\linewidth]{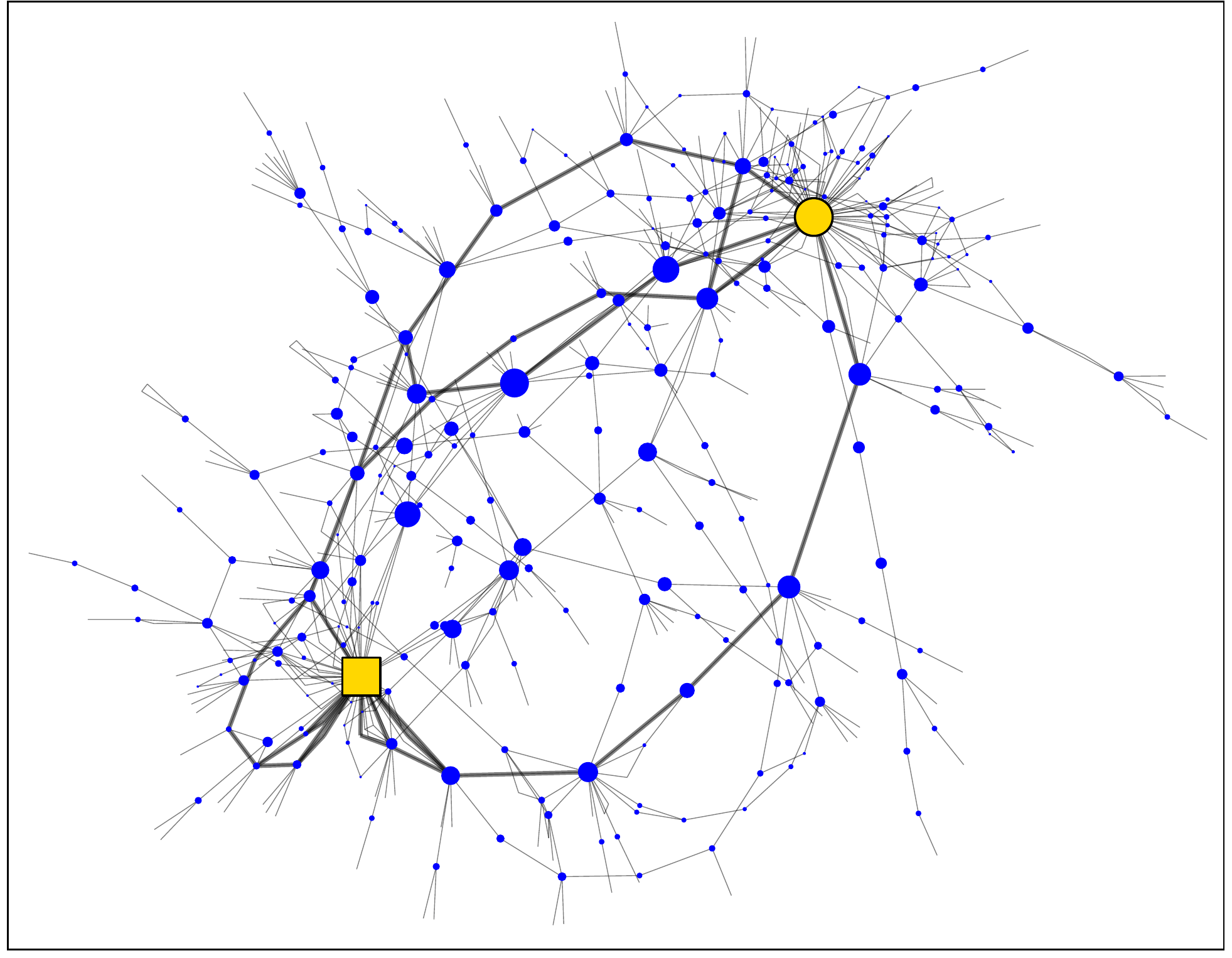}
a) Betweenness centrality 
\end{minipage}
\begin{minipage}{0.48\linewidth}
\centering
\includegraphics[width=\linewidth]{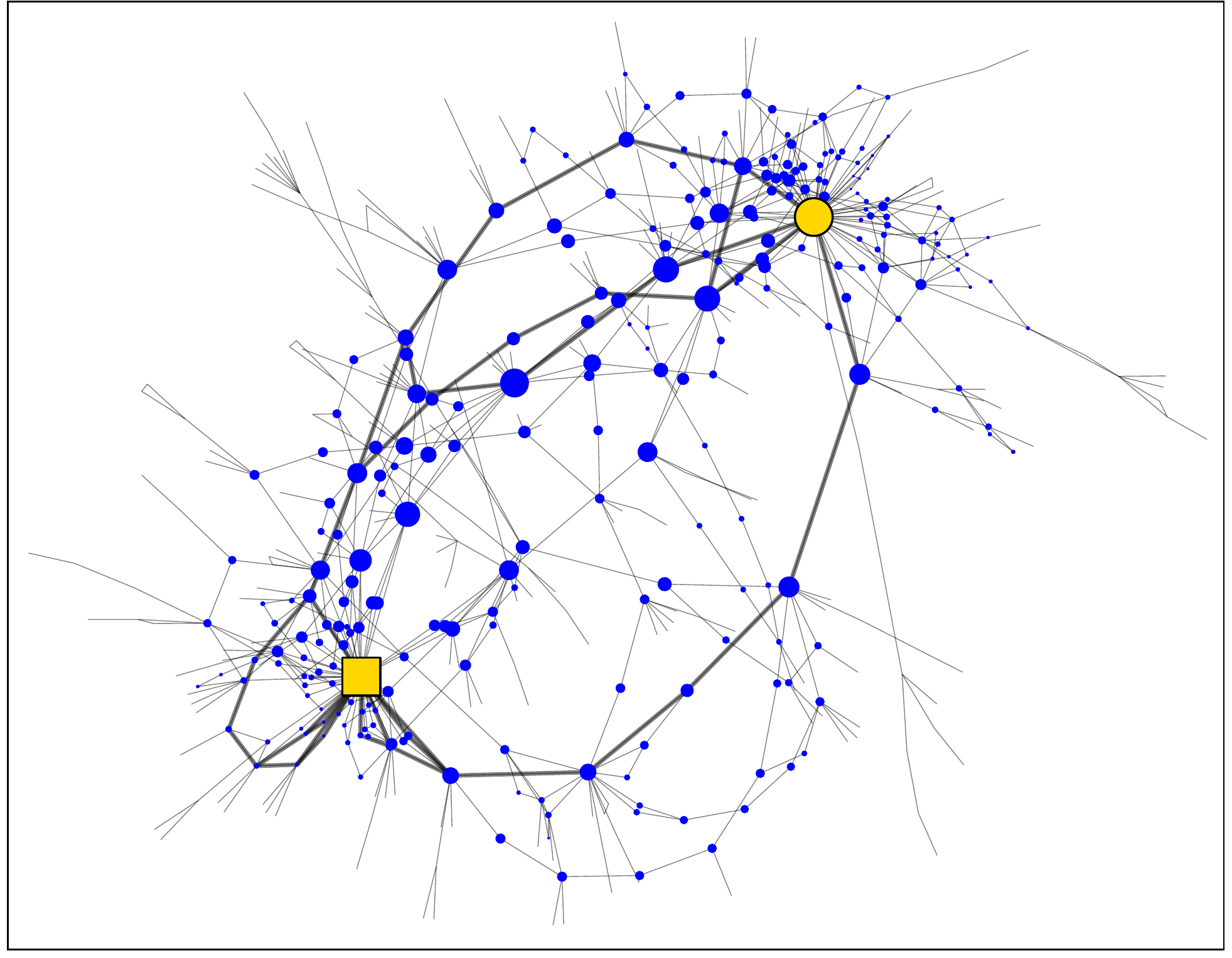}
b) Source-to-target current flow 
\end{minipage}

\vspace{10pt}

\begin{minipage}{0.48\linewidth}
\centering
\includegraphics[width=\linewidth]{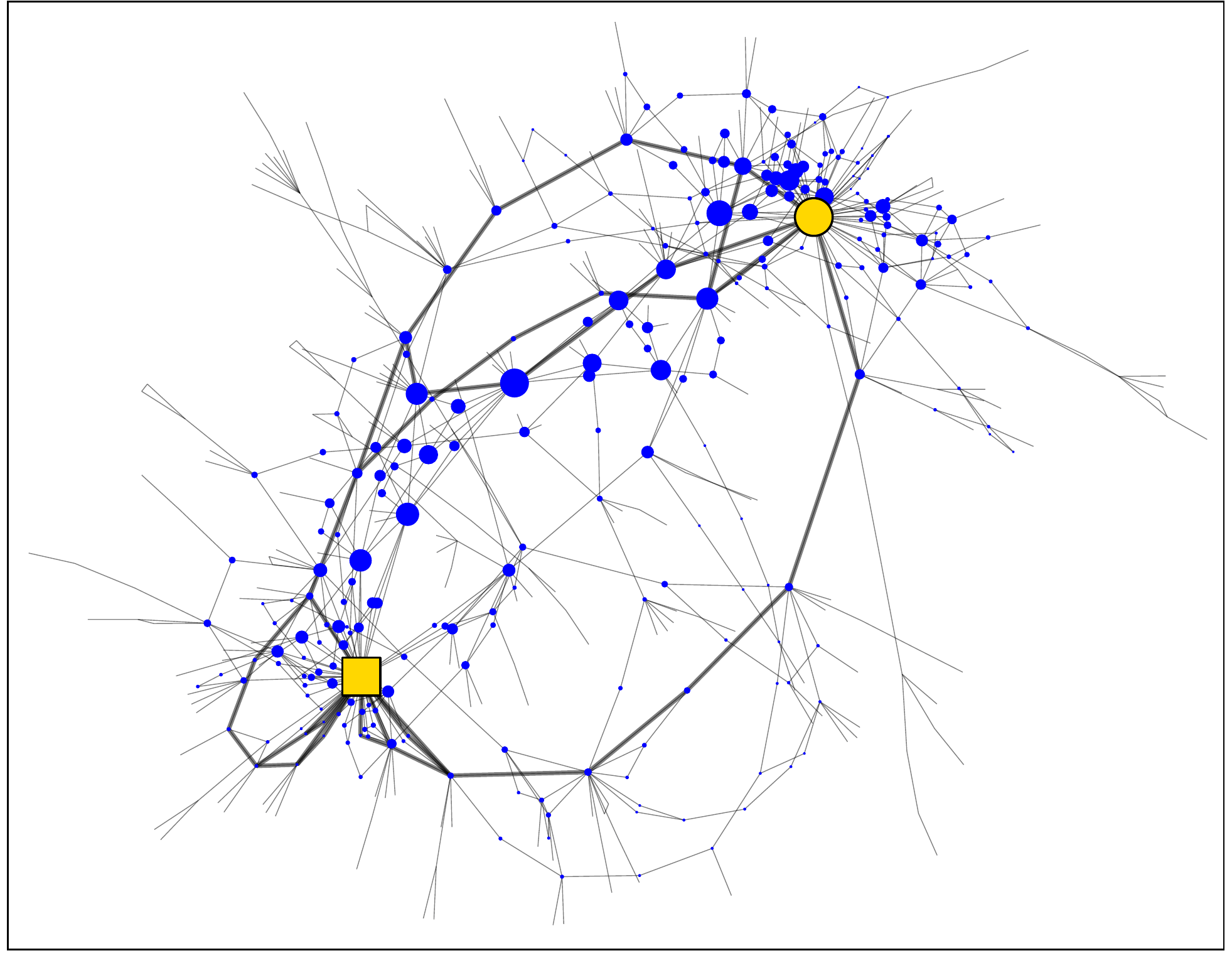}
c) Source-to-target simple paths 
\end{minipage}
\begin{minipage}{0.48\linewidth}
\centering
\includegraphics[width=\linewidth]{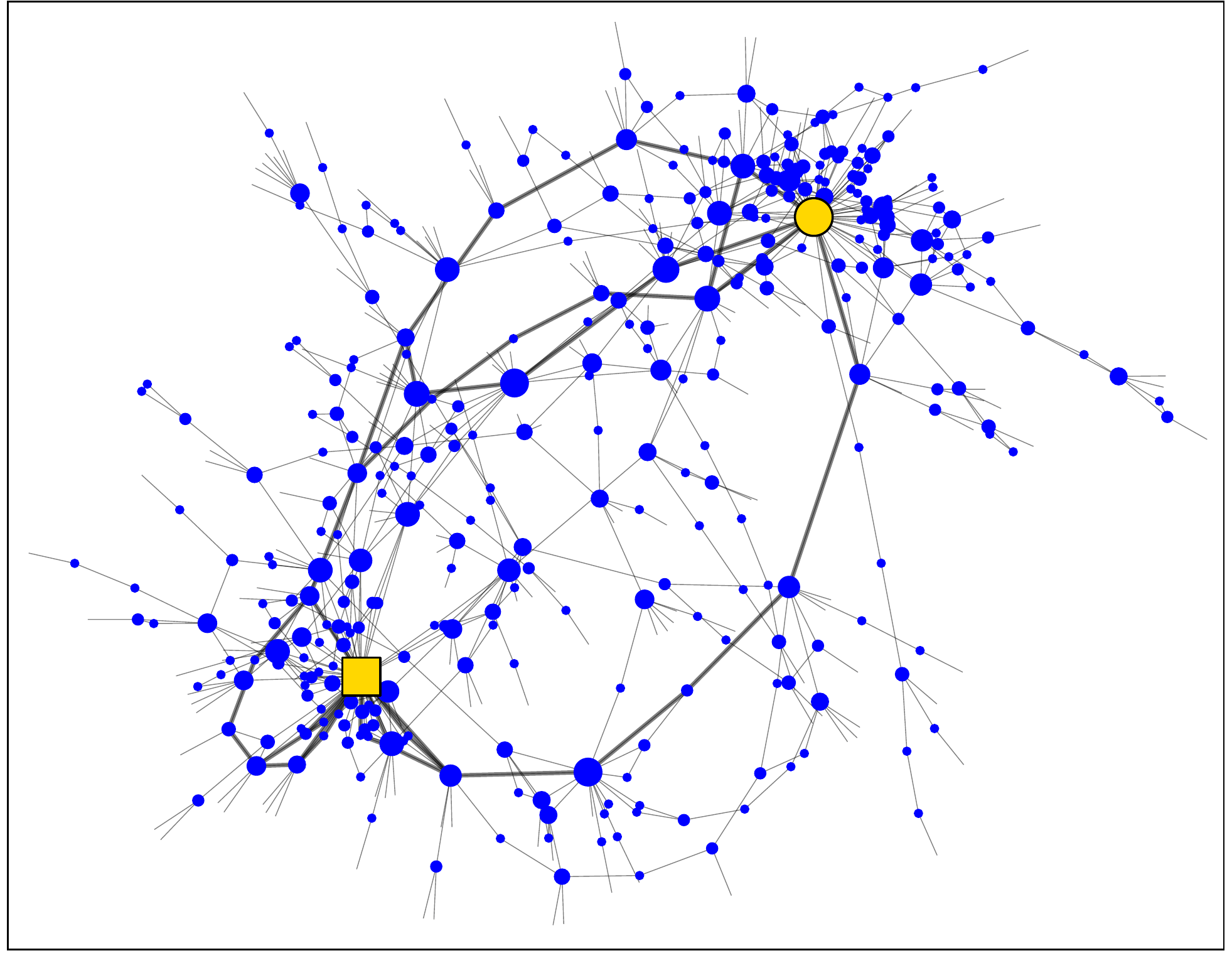}
d) Degree centrality 
\end{minipage}

\vspace{10pt}

\begin{minipage}{0.48\linewidth}
\centering
\includegraphics[width=\linewidth]{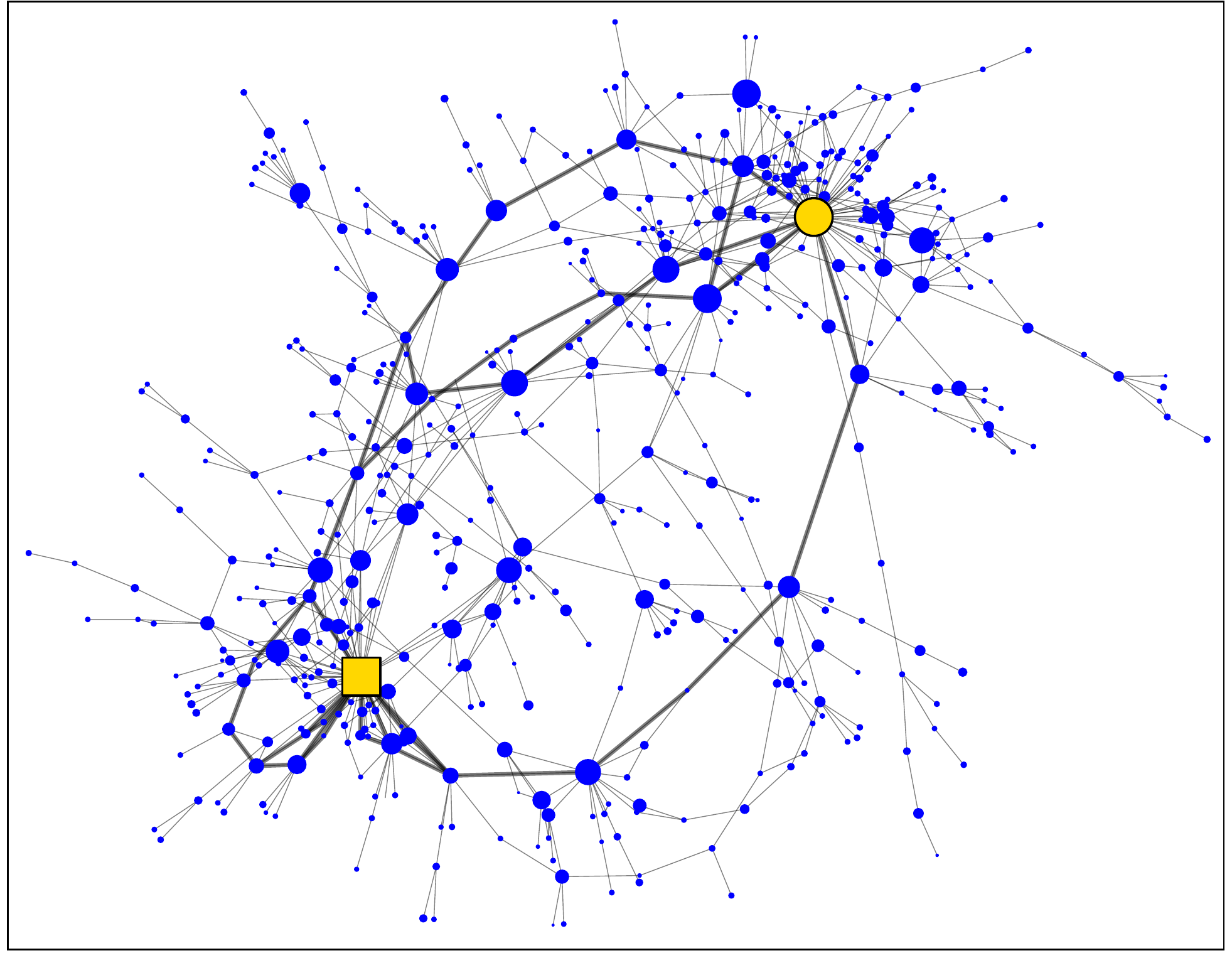}
e) Projected volume 
\end{minipage}
\begin{minipage}{0.48\linewidth}
\centering
\includegraphics[width=\linewidth]{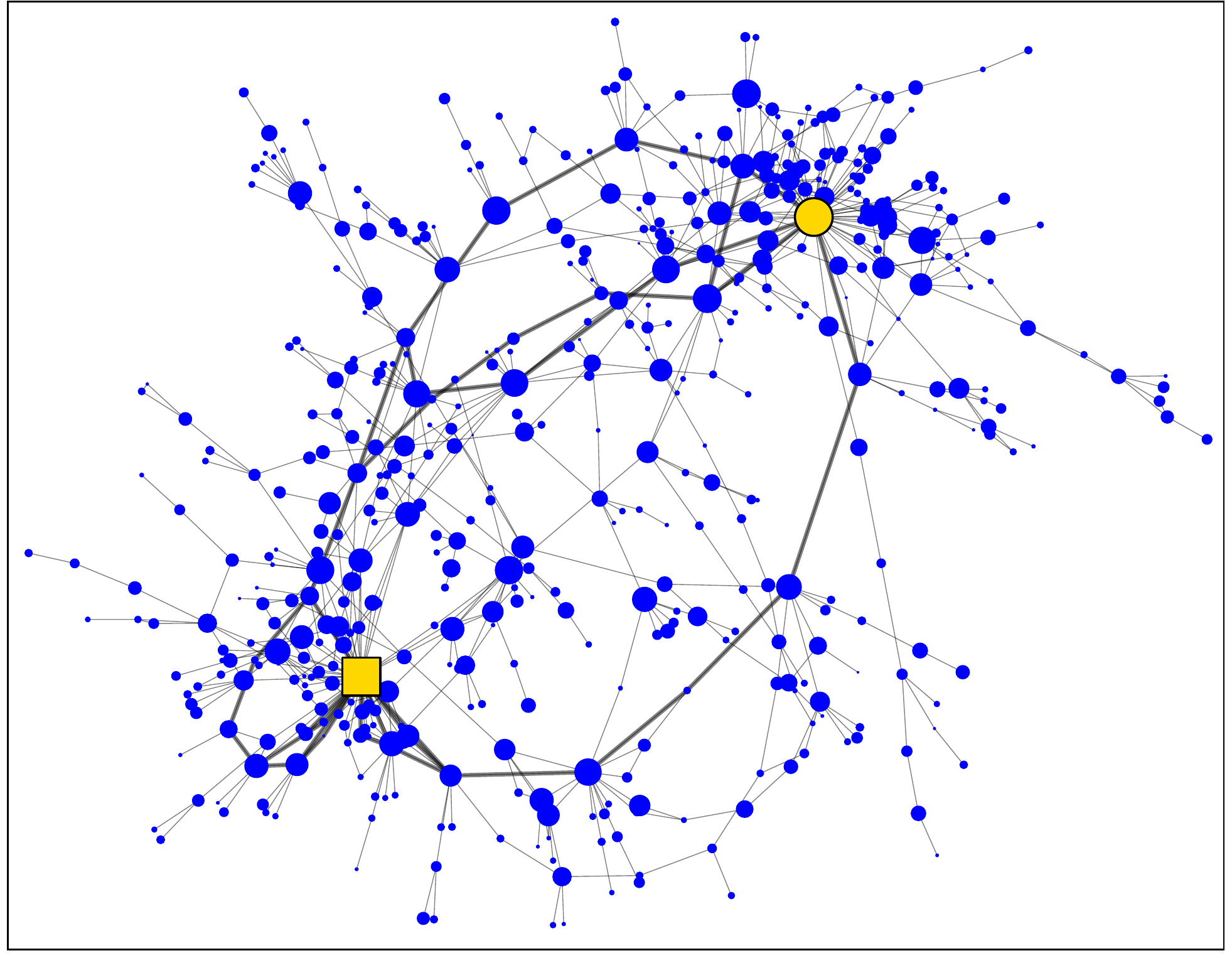}
f) Permeability 
\end{minipage}

\caption{Visualization of a graph derived from a random DFN as shown in Figure~\ref{fig:dfn}.  Blue circles represent normalized feature values using six different features, in panels a) through f). Yellow square denotes source, yellow circle denotes target.  Heavy lines represent particle backbone paths in the graphs. Note varying extent of correlation between particle backbone and associated feature strength.}\label{fig:feature_viz}

\end{figure}

The centrality of a node in a graph describes its importance to transport across the network.
Motivated by recent studies suggesting centrality measures that can help identify regions important in conducting flow and tansport~\cite{Santiago2014,Santiago2016},
we consider four such quantities as node features.
Three are global topological measures, quantifying how frequently paths through the graph include a given node.  
One is a local topological measure, giving the number of immediate neighbors of a node.
Graph properties are computed using the {\sc NetworkX} graph software package~\cite{hagberg-2008-exploring}.  
The topological features are supplemented with two physical (geometric) features; fracture volume projected along the main flow axis (from inlet plane to outlet plane) and fracture permeability.
Figure~\ref{fig:feature_viz} provides a visualization of a graph derived from the random DFN shown in Figure~\ref{fig:dfn}.  Blue circles represent normalized feature values using these six different features, in panels a) through f). The yellow square and circle denote the source and target.  Heavy lines represent particle backbone paths in the graphs.

\subsubsection{Global topological features}

\begin{itemize}
\item
The \emph{betweenness centrality}~\cite{Anthonisse1971,Freeman1977} of a node (Figure \ref{fig:feature_viz}a) reflects the extent to which that node can control communication on a network.  Consider a geodesic path (path with fewest possible edges) connecting a node $u$ and a node $v$ on a graph.  In general, there may be more than one such path: let $\sigma_{uv}$ denote the number of them.  Furthermore, let $\sigma_{uv}(i)$ denote the number of such paths that pass through node $i$.  We then define, for node $i$,

\begin{equation}
\label{eq:betweeness}
\text{Betweenness centrality} = \frac{1}{(n-1)(n-2)}\sum_{\substack{u,v = 1 \\ u\neq i\neq v }}^n \frac{\sigma_{uv}(i)}{\sigma_{uv}},
\end{equation}
\noindent
where the leading factor normalizes the quantity so that it can be compared across graphs of different sizes $n$.
Figure \ref{fig:feature_viz}a confirms that many backbone nodes do indeed have high betweenness values. At the same time, certain paths through the network that are not part of the backbone also show high values for this feature, reflecting that betweenness centrality considers \emph{all} paths in the graphs, and not only those from source to target. 

\item
\emph{Source-to-target current flow} (Figure \ref{fig:feature_viz}b) is a centrality measure based on an electrical current model~\cite{Brandes2005}, and assumes a given source and target.  Imagine that one unit of current is injected into the network at the source, one unit is extracted at the target, and every edge has unit resistance.  Then, the current-flow centrality at a node is equal to the current passing through it.  This is given by Kirchhoff's laws, or alternatively in terms of the graph Laplacian matrix $\mathbf{L}=\mathbf{D}-\mathbf{A}$, where $\mathbf{A}$ is the adjacency matrix for the graph and $\mathbf{D}$ is a (diagonal) matrix specifying node degree: $D_{ii}=\sum_j A_{ij}$.  Letting $\mathbf{L}^+$ denote the Moore-Penrose pseudoinverse of $\mathbf{L}$, $s$ the source node, and $t$ the target node, then for node $i$ we define

\begin{equation}
\label{eq:currentflowbetweeness}
\text{Current flow } = 
\sum_{j=1}^n A_{ij} \bigl| \bigl(L_{is}^{+} - L_{js}^{+}\bigr) - \bigl(L_{it}^{+} - L_{jt}^{+}\bigr) \bigr|.
\end{equation}

\noindent
Current-flow centrality is also known as random-walk centrality~\cite{Newman2005}, since the same quantity measures how often a random walk from $s$ to $t$ passes through $i$.
Unlike betweenness centrality, the current-flow centrality is zero on any branch of the graph outside of the central core. We therefore expect high current flow values to correlate with nodes that have large influence on source-to-target transport.

\item
\emph{Source-to-target simple paths} (Figure \ref{fig:feature_viz}c) is a centrality measure that counts simple (non-backtracking) paths crossing the graph from source $s$ to target $t$. Let $\pi_{st}$ denote the number of such paths, and $\pi_{st}(i)$ denote the number of those passing through node $i$.  We then define, for node $i$,

\begin{equation}
\label{eq:source-to-target}
\text{Simple paths} = \frac{\pi_{st}(i)}{\pi_{st}}, 
\end{equation}

\noindent
where normalization by $\pi_{st}$ allows comparing values of simple path centrality across different graphs. 
Due to the exponential proliferation of possible paths, we limit our search to paths with 15 nodes or less.  Beyond 15, the effect on source-to-target simple path centrality  is negligible.
Figure \ref{fig:feature_viz}c illustrates that nodes with high simple path centrality are more likely to lie on backbone paths than are nodes with high betweenness centrality in Figure \ref{fig:feature_viz}a. However, simple path centrality also fails to identify one isolated backbone path that is disjoint from the others. 

\end{itemize}

\subsubsection{Local topological feature}
\begin{itemize}
\item
\emph{Degree centrality} (Figure \ref{fig:feature_viz}d) is a normalized measure of the number of edges touching a node.  For node $i$,
\begin{equation}
\label{eq:degree_cent}
\text{Degree centrality} = \frac{1}{n-1} \sum_{j=1}^n A_{ij}.
\end{equation}
\noindent
Nodes with high degree centrality tend to be concentrated in the core of the network.  Conversely, nodes with low degree centrality are often in the periphery or on branches that cannot possibly conduct significant flow and transport. Physically, degree centrality of a fracture measures the number of other fractures that intersect with it.
\end{itemize}

\subsubsection{Physical features}

We supplement the four topological features with two features describing physical properties of fractures.

\begin{itemize}
\item
\emph{Projected volume} (Figure \ref{fig:feature_viz}e) measures the component of a fracture's volume oriented along the direction of flow from inlet to outlet plane.  
Let fracture $i$ have volume $V_i$ and orientation vector $\mathbf{O}_i$ (unit vector normal to the fracture plane).  Taking flow to be oriented along the $x$-axis, the projected volume is expressed by the projection of $\mathbf{O}_i$ onto the $yz$-plane: 

\begin{equation}
\label{eq:volprojection}
\text{Projected volume} = V_i \sqrt{(O_i)_y^2 + (O_i)_z^2}.
\end{equation}

Figure \ref{fig:feature_viz}e shows similarities between this feature and degree centrality, but also some fractures where one feature correlates more closely with the backbone than the other.

\item
\emph{Permeability} (Figure \ref{fig:feature_viz}f) measures how easily a porous medium allows flow to pass therein.  Given the aperture size $b_i$ of fracture $i$, the permeability is expressed as

\begin{equation}
\label{eq:highDegreek=1}
\text{Permeability} = \frac{b_i^2}{12} 
\end{equation}

The permeability of a fracture, which is nonlinearly related to its volume, is a measure of its transport capacity.  As illustrated in Figure \ref{fig:feature_viz}, it displays similarities to both degree centrality and projected volume, with backbone fractures almost systematically having high permeability values (but the converse holding less consistently).
\end{itemize}

\subsection{Correlation of feature values}

\begin{figure}[t]
\begin{center}
\includegraphics[scale=.45]{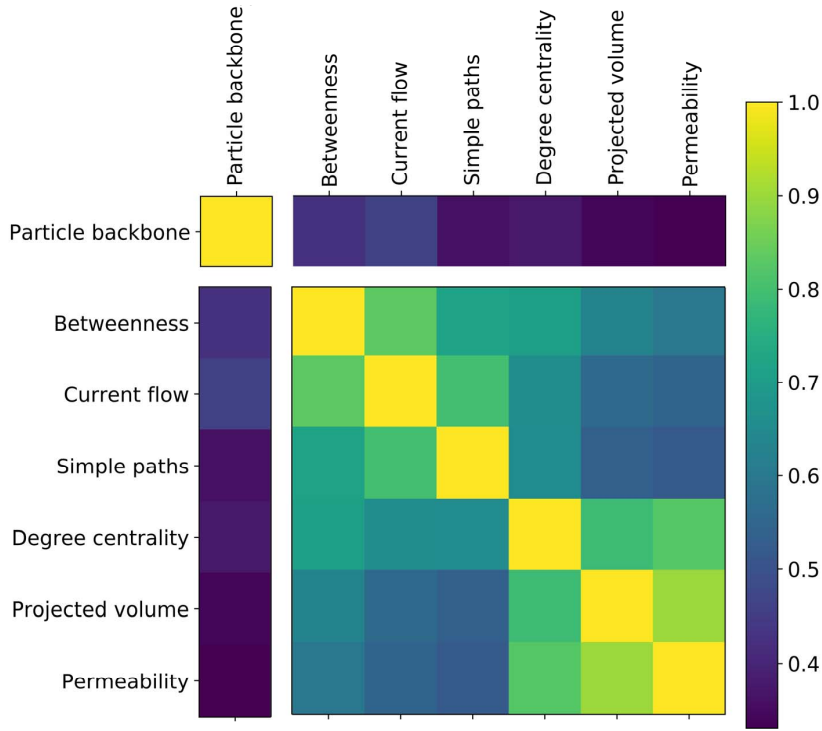}
\centering
\caption{Heat map displaying correlations among the particle backbone and the six features used.}
\label{fig:correlation}
\end{center}
\end{figure}

As is seen in Figure \ref{fig:feature_viz}, the feature values vary widely from one node to another, in ways that we aim to manipulate in order to generate subnetworks. Figure \ref{fig:correlation} shows correlation coefficients for pairs that include the particle backbone and the six features that we have chosen.  That there are non-negligible correlations between the backbone and these features suggests that they are relevant ones for classification, although clearly no single feature is sufficient in itself.  The correlation coefficients indicate that features tend to cluster naturally into the three categories above.  The first three features, which are the global topological ones (betweenness, current flow, and simple paths), have significant mutual correlations among then.  The same is true for the physical features (projected volume and permeability), which also exhibit some clustering with the related local topological feature (degree centrality).  The latter correlations are consistent with our feature definitions above.

While we also considered additional centrality measures studied in the literature~\cite{Santiago2014,Santiago2016}, such as closeness, eccentricity, and eigenvector centrality, we found that those exhibited much weaker correlations with the particle backbone.  The choice of the six features above was motivated through cross-validation tests, where adding centrality measures did not improve classification performance while removing any of them degraded performance.

\section{Classification Methods}
\label{S:methods}

In this section, we explain how we evaluate classification performance, briefly describe our two machine learning algorithms (random forest and support vector machines), and our process for parameter selection. Detailed descriptions of the methods are provided in the appendix.
Both algorithms are general-purpose supervised learning methods suitable for geometric as well as non-geometric features. 
Given a set of features and class assignment for some observations, supervised learning algorithms try to ``learn'' the underlying function that maps features to classes. Those observations are the training set. Once learned, the function can then be used to classify new observations. In our study, we use as observations the nodes (fractures) from 80 graphs as a training set. We then test the function using nodes from 20 graphs as a test set.
 Both algorithms are implemented using the \emph{scikit-learn} machine learning package in python, with the functions \emph{RandomForestClassifier} and \emph{SVC}.


\subsection{Performance measures}

There are a number of challenges in evaluating classification performance.  Our problem has a large class imbalance: only about 7 percent of nodes in the training set are in the particle backbone.  A classifier could simply assign all nodes to the non-backbone class, and still achieve an overall accuracy of 93 percent.  Moreover, our approach departs from more conventional machine learning methodology in that our ultimate objective is not necessarily a perfect recovery of the backbone.  We train on the particle backbone in order to identify a subset of fractures that share its characteristics, thereby reducing the full network to a subnetwork with analogous flow behavior.  We then validate the flow behavior using the breakthrough curve (BTC), which describes the distribution of times for particles to pass through the network.  For these reasons, performance measures must be interpreted with care.

For straightforward backbone prediction, we may define a \emph{positive} classification of a node as being an assignment to the backbone class, and a \emph{negative} classification as being an assignment to the non-backbone class.  True positives (TP) and true negatives (TN) represent nodes whose backbone/non-backbone assignment matches that of the labeled data.  False positives (FP) and false negatives (FN) represent nodes whose backbone/non-backbone assignment is opposite that of the labeled data.  One measure of success is the TP rate.  \emph{Precision} ($p$) and \emph{recall} ($r$) represent two kinds of TP rates:
\begin{equation}
\label{eq:precision}
p = \frac{TP}{TP + FP}
\end{equation}

\begin{equation}
\label{eq:recall}
r = \frac{TP}{TP + FN}
\end{equation}
Precision is the number of true positives over the total number that we classify as positive, whereas recall is the number of true positives over the total number of actual positives.  These values give an understanding of how reliable (precision) and complete (recall) our results are.

\begin{figure}[!b]
	\centering
	\includegraphics[width=0.8\linewidth]{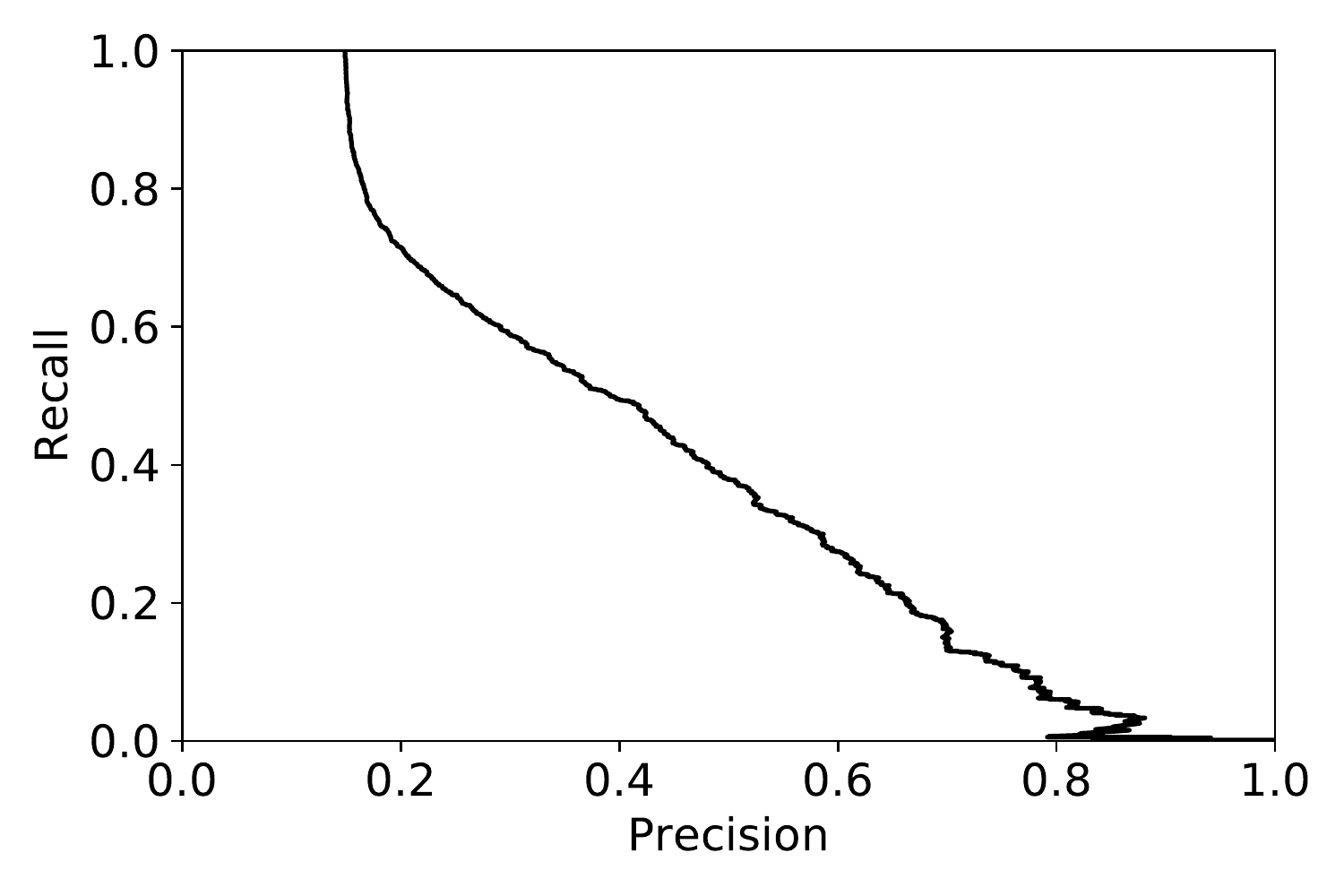}
	\caption{Precision/recall curve for current-flow thresholding as a classifier.  As threshold value increase, classifier moves from perfect recall to perfect precision.  Ideally, precision/recall curve would touch upper-right corner for strongest classifier performance.}
	\label{fig:prcurve_currentflow}
\end{figure}

There is a tradeoff between precision and recall.  This can be seen in the behavior of one of the simplest possible forms of classification: thresholding according to a single feature.  Consider a classifier that labels as backbone all fractures with nonzero current flow.  The process would resemble the dead-end fracture chain removal method common in the hydrology literature, but more extreme in that it would eliminate all dead-end subnetworks.  This gives perfect (100\%) recall, since all fractures in the particle backbone necessarily have nonzero current flow, and 15\% precision, as it reduces the network to approximately half of its original size (about 7\% of which were TP).  Now imagine increasing the threshold, so as to lower the number of positive assignments.  This will reduce FP, thereby increasing the precision.  However, it will also increase FN, reducing the recall.  In this way, we can travel along a precision/recall curve, shown in Figure~\ref{fig:prcurve_currentflow}, that has perfect recall as one extreme and perfect precision as the other.  If there existed a threshold value at which the classifier recovered the class labels perfectly, the precision/recall curve would touch the upper-right corner of the figure: 100\% precision and 100\% recall.  One typically wants classifiers that come as close to that ideal as possible.

In order to quantify the tradeoff shown in Figure~\ref{fig:prcurve_currentflow}, one may introduce a utility function
\begin{equation}
\label{eq:utility}
u_\alpha({\mathbf q}) = \alpha p({\mathbf q}) + (1-\alpha) r({\mathbf q}),
\end{equation}
where $\alpha\in[0,1]$ specifies the relative weight of precision vs.\ recall, $\mathbf{q}$ denotes a vector of hyperparameter values used for tuning the classifier, and $p({\mathbf q})$ and $r({\mathbf q})$ denote the precision and recall obtained by the classifier for those parameter values.  Thus, $u_\alpha({\mathbf q})$ is determined purely by recall when $\alpha=0$, and purely by precision when $\alpha=1$.  In the example above, ${\mathbf q}$ would be a scalar quantity $q$ denoting the current-flow threshold value.  In general, for a given weight $\alpha$, we may find the hyperparameter values $\mathbf{q}$ that maximize $u_\alpha({\mathbf q})$.  The most straightforward algorithm for doing so is grid search cross-validation, which performs an exhaustive search over a given range of $\mathbf{q}$ values, evaluating $p({\mathbf q})$ and $r({\mathbf q})$ based on subsampling of the training data.  This procedure avoids overfitting that would occur from validation using only the test data.

While the tradeoff between precision and recall is related to the tradeoff between network reduction and accuracy, it is not identical.  Network reduction is measured by the ratio
\begin{equation}
\label{eq:fractures_remaining}
\text{Fractures remaining} = \frac{TP+FP}{n} = \frac{r}{p}\,\frac{TP+FN}{n} = \frac{r}{p}\,\beta,
\end{equation}
where $n$ is the number of fractures in the full network and $\beta$ is the proportion of fractures that are in the particle backbone.  Low recall and high precision therefore yield small subnetworks.  Accuracy is measured by the agreement between the BTC of the subnetwork and of the full network.  High accuracy correlates with high recall: we train our classifier on the particle backbone because it is a valid network reduction from the perspective of characterizing where the majority of flow, and thus transport, occurs.  But it is only one of many valid reductions. Ideally, one could optimize accuracy by computing the BTC for the subnetwork predicted with each choice of hyperparameters $\mathbf{q}$ in our grid search, and comparing with the full network's BTC. Unfortunately, such a framework is computationally infeasible, since meshing a DFN, solving for flow and transport requires tens of minutes of wall clock time for each set of $\mathbf{q}$ values. Consequently, in place of high accuracy, we aim for high (though not necessarily perfect) recall.  Precision is less essential: false positives increase the size of the predicted subnetwork, but for a small backbone ($\beta\approx 0.07$), even low precision allows for significant network reduction.

\subsection{Random forest}\label{sec:RF}

A random forest~\cite{ho_rf_1995,ho_rf_1998} is constructed by sampling from the training set with replacement, so that some data points may be sampled multiple times and others not at all. Those data points that are sampled are used to generate a large collection of decision trees, each of which outputs a classification based on feature values. Those data points that are not sampled are run through the decisions trees. 
A test data point is then classified by having each tree ``vote'' on its class.  This leads not only to a predicted classification, but also to a measure of certainty (the fraction of trees that voted for it) as well as to an estimate of the importance of each feature for the class assignment~\cite{Breiman2001}.
That final estimate is particularly useful when the features consist of quantities that measure different aspects of node centrality. 
Further discussion of the random forest method is provided in the appendix.

In order to identify the hyperparameters of random forest that affect our results most significantly, we use the grid search cross-validation method described above, implemented with the \emph{GridSearchCV} function in \emph{scikit-learn}.
We aim for high recall (low $\alpha$, in Eq.~(\ref{eq:utility})),  and find the greatest sensitivity to a hyperparameter that sets the minimal number of samples in a leaf node, to limit how much a decision tree branches.  This is the sole hyperparameter for our classifier, so the vector ${\mathbf q}$ in Eq.~(\ref{eq:utility}) reduces to a scalar quantity $q$.  
Adjusting its value prevents overfitting, which in the context of unbalanced classes could cause practically none of the feature space to be assigned to the minority class~\cite{Chen2004}.
\subsection{Support vector machines}\label{sec:SVM}

Support vector machines (SVM) separate high-dimensional data points into two classes by finding an appropriate hyperplane.  Based on the generalized portrait algorithm~\cite{Vapnik1963} and subsequent developments in statistical learning theory~\cite{Vapnik1974}, the current version of SVM~\cite{cortes_svm_1995} uses kernel methods~\cite{Boser1992} to generalize linear classifiers to nonlinear ones.  These are discussed further in the appendix.  Kernel methods enable SVM to perform well when certain feature variables are highly correlated or even unimportant to the class assignment~\cite{James2013}, and help prevent overfitting.  We therefore enlarge our feature space for SVM by ranking the values of each feature on the nodes of a given graph.  For a given feature, if $n$ nodes have feature values $f_1,\dots,f_n$, then we define ranked features $\hat{f}_1,\dots,\hat{f}_n$ whose values are given by the order statistics of $f$, i.e.,
\begin{eqnarray}
\hat{f}_i = 1 & \text{if} & f_i = \min\{f_1,\dots,f_n\}, \\
\hat{f}_i = n & \text{if} & f_i = \max\{f_1,\dots,f_n\},
\end{eqnarray}
and generally, the ranked feature $\hat{f}_i=k$ if the ``raw'' feature $f_i$ is equal to the $k$th order statistic $f_{(k)}$.  We supplement the collection of six raw features discussed in Section~\ref{S:features} with the six corresponding ranked features, resulting in a total of twelve features.

As with random forest, we use grid search cross-validation to identify and optimize crucial hyperparameters in SVM.  We find the most important of these to be the penalty parameter,
a regularization coefficient that controls the strictness of the decision boundary.  When the penalty is large, SVM imposes a hard (rough) boundary in the training data, at the risk of overfitting. When the penalty is small, SVM allows a smoother boundary and more misclassification among the training data.
Because of our class imbalance, we assign different penalty values for each class, so that the classifier more strictly bounds the (minority) backbone nodes than the (majority) non-backbone nodes~\cite{Osuna1997}.  These quantities form the two hyperparameters ($\mathbf q$ in Eq.~\ref{eq:utility}) for our classifier.  In this way, we can simultaneously prevent overfitting the majority class and ``overlooking'' points in the minority class.  By adjusting the balance of penalty values, we control how likely the classifier is to assign a node to the backbone.

\section{Results}
\label{S:results}

We used a collection of 100 graphs.  80 were chosen as training data, and 20 were chosen as test data.  We illustrate certain results, including breakthrough curves, on the DFN shown in Figure~\ref{fig:dfn}.  Other results are based on the entire test set, which consists of a total of 9238 fractures, 651 of which (7.0\%) are in the particle backbone and 8587 of which (93\%) are not.  The total computation time to train both RF and SVM was on the order of a minute, negligible compared to the time to extract the particle backbone needed for training.  Once trained, the classifier ran on each test graph in seconds.

\subsection{Classifiers}
\label{sec:importance}

\begin{table}[!b]
\begin{center}
	\begin{tabular}{c|c|c|c}
 	Classifier & Precision & Recall & Fractures remaining \\ \hline
	RF(1400) & 18\% & 90\% & 36\%  \\
	RF(30) & 26\% & 75\% & 21\%  \\
	RF(15) & 30\% & 65\% & 15\%  \\
	RF(1) & 58\% & 20\% & 2.5\% 
	\end{tabular}
\caption{Random forest classifiers labeled by the \emph{min\_samples\_leaf} parameter value, controlling how much a decision tree can branch.  Percentages for precision, recall, and  fractures remaining in network are calculated over all 20 graphs in the test set.}
\label{table:rf-models}
\end{center}
\end{table}

We implemented random forest using the \emph{RandomForestClassifier} function in \emph{scikit-learn}, on the six features described in Section~\ref{S:features}. Based on cross-validation with the function \emph{GridSearchCV}, we found default hyperparameter values to be sufficient for achieving high recall, except as follows: 250 trees (\emph{n\_estimators=250}); best split determined by binary logarithm of number of features (\emph{max\_features=log2}); information gain as the quality measure for a split (\emph{criterion=`entropy'}); voting weights inversely proportional to class frequency (\emph{class\_weight=`balanced\_subsample'}). We varied the minimal number of training samples in a leaf node (\emph{min\_samples\_leaf}) to adjust the tradeoff between recall and precision.  Table~\ref{table:rf-models} shows the results for a sample of four different hyperparameter values; one can also explicitly find the value that maximizes the utility function $u_\alpha(q)$ for a sample of different $\alpha$ values, using \emph{GridSearchCV}.
Since the full particle backbone accounts for only 7\% of the fractures in the test set, we see that even classifiers with relatively low precision can reduce the network significantly.

\begin{figure}[t]
	\centering
	\includegraphics[width=0.8\linewidth]{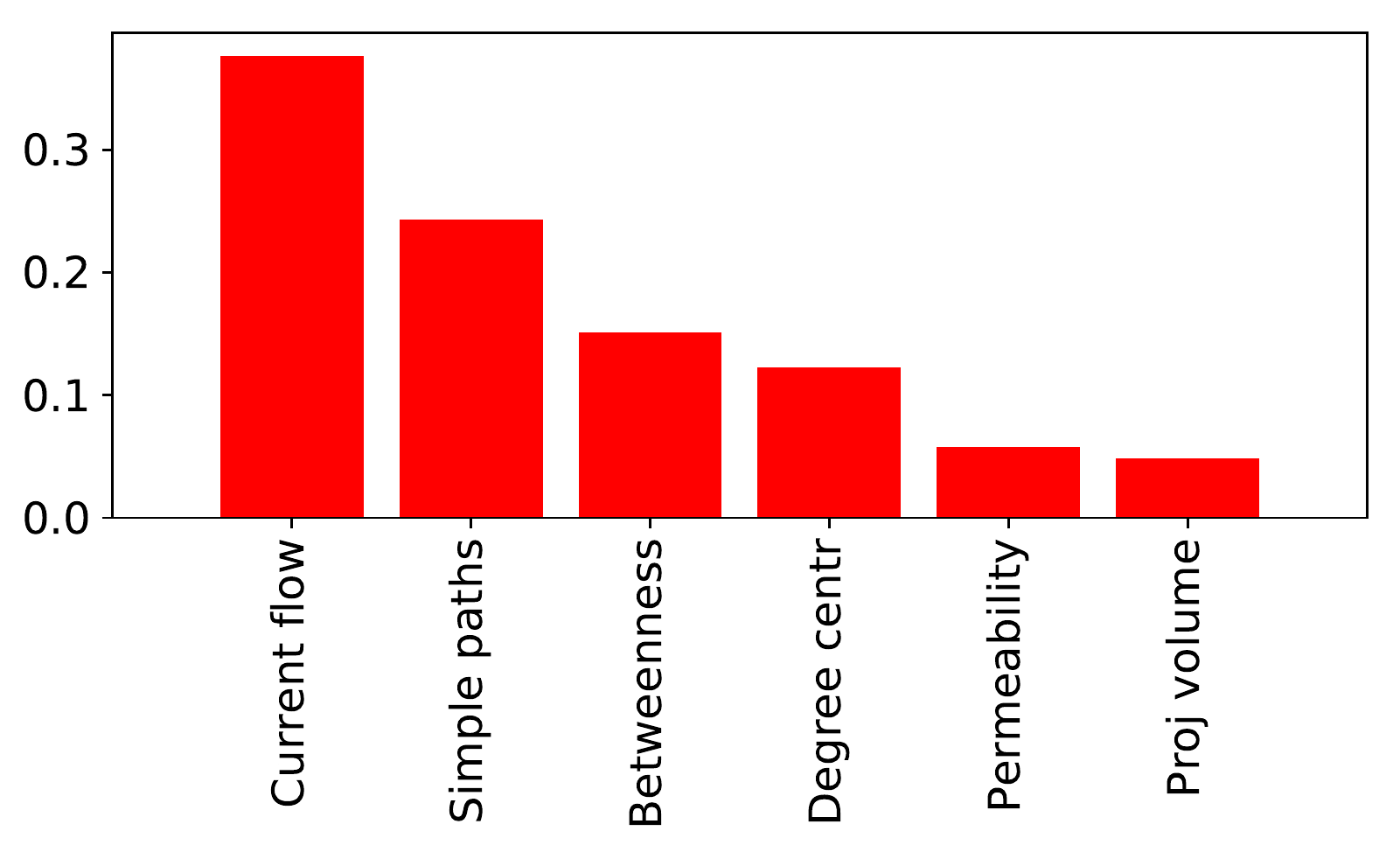}
 	\caption{Relative importances of features based on training data for random forest.}
	\label{fig:importance}
\end{figure}

Random forest also provides a quantitative estimate of the relative importance of each of the six features described in Section~\ref{S:features}, based on how often a tree votes for it.  Using the RF(30) model on our 80 training graphs, we find the feature importances shown in Figure~\ref{fig:importance}.  
The source-to-target current flow, source-to-target simple paths, and betweenness centralities are the most important features, followed by node degree, and followed finally by permeability and projected volume.  Thus, as with the feature correlations shown in Figure~\ref{fig:correlation}, the feature importances cluster into three natural groups.  Global topological features have the greatest importance, local topological features have significant but lower importance, and physical features play only a small role in classification.  In contrast with SVM, the performance of random forest does not benefit from using additional features such as ranked features.  The inherent bootstrapping of random forest enables strong classification performance even with a relatively limited number of features.

We implemented SVM using the \emph{SVC} function in \emph{scikit-learn}, on the twelve features made up of the six raw features described in Section~\ref{S:features} along with their ranked counterparts.
We chose penalty value pairs (\emph{class\_weight}, often called $C$ in the literature~\cite{cortes_svm_1995,Osuna1997}) for the backbone and non-backbone class, which we adjusted in order to vary precision and recall. Table~\ref{table:svm-models} shows results for a sample of four different pairs of values.  Similarly to RF, one could find value pairs ${\mathbf q}$ that maximize the utility function $u_\alpha({\mathbf q})$ for a sample of different $\alpha$ values, by using the \emph{GridSearchCV} cross-validation function.
All other parameters were set to their default values, which include a radial kernel (closed decision boundary).

\begin{table}[ht]
\begin{center}
\begin{tabular}{c|c|c|c}
 	Classifier & Precision & Recall & Fractures remaining \\ \hline
	SVM(0.90,0.054) & 17\% & 96\% & 39\%  \\
	SVM(0.90,0.063) & 19\% & 90\% & 34\%  \\
	SVM(0.70,0.070) & 23\% & 78\% & 24\%  \\
	SVM(0.70,0.190) & 44\% & 46\% & 7.3\% 
\end{tabular}
\caption{SVM classifiers labeled by the \emph{class\_weight} parameter pair values, specifying misclassification penalties for the backbone and non-backbone classes. Percentages for precision, recall, and fractures remaining in network are calculated over all 20 graphs in the test set.}
\label{table:svm-models}
\end{center}
\end{table}

\subsection{Validation}
\label{sec:validation}

\begin{figure}[!b] \centerline{
 \begin{tabular}{ccc}
 (a)   \\
 \includegraphics[width=0.7\textwidth]{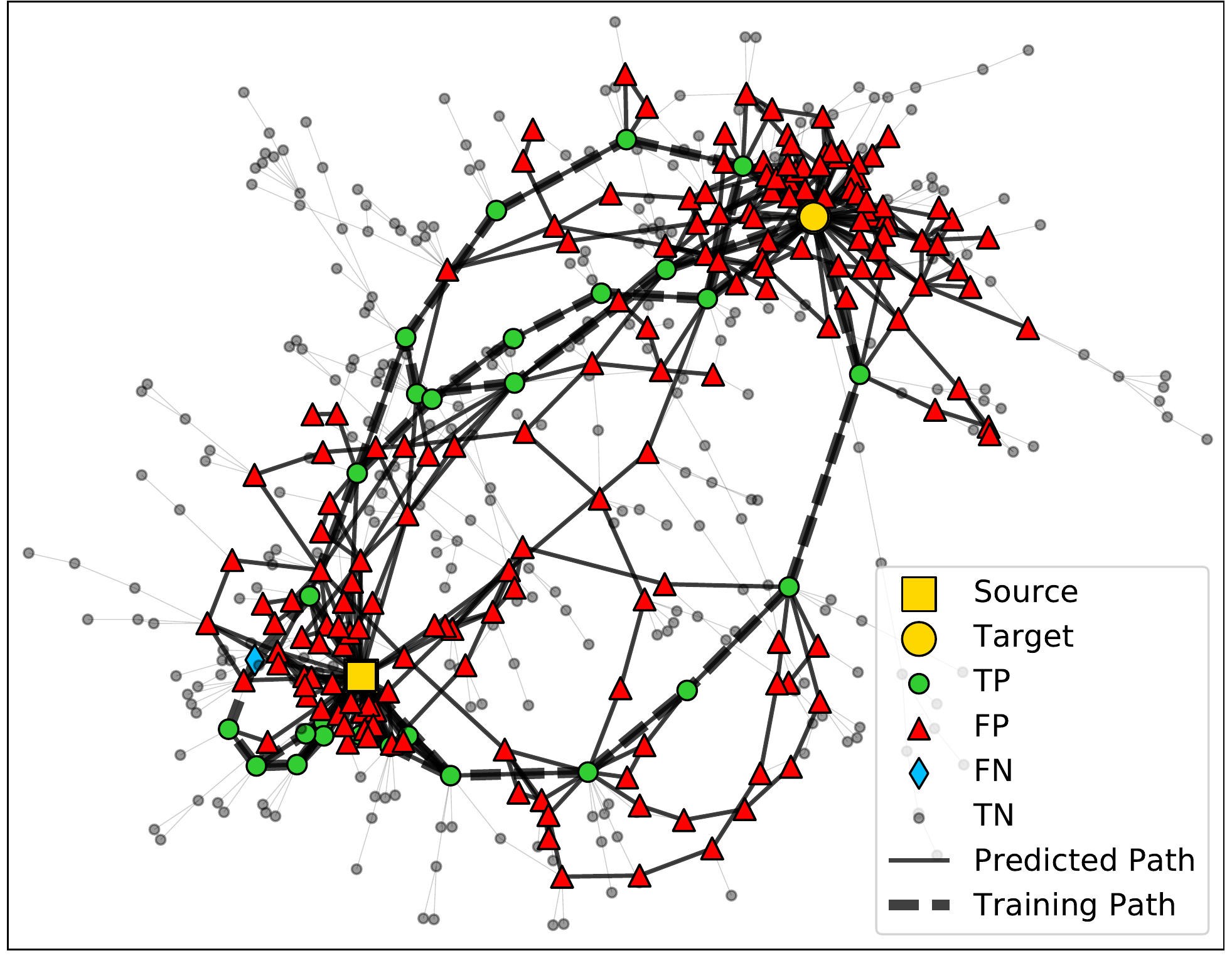} \\
 \ \\
 (b)   \\
  \includegraphics[width=0.7\textwidth]{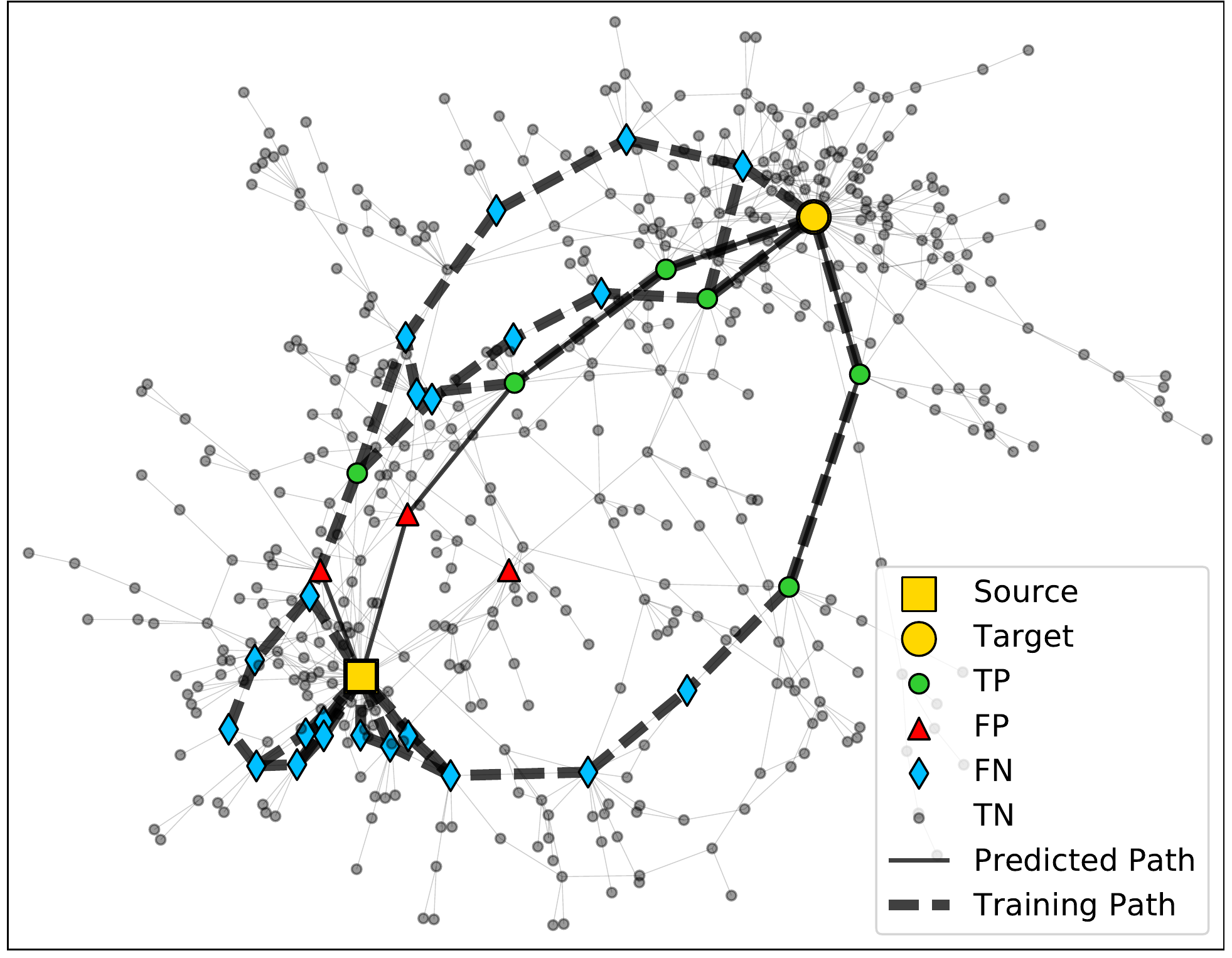} \\
 \end{tabular}}
\caption{\label{fig:18p-91r} Extreme cases of classification results: (a) SVM(0.90,0.054) with high recall and low precision (40\% of network remaining), showing many false positives (FP) and relatively few false negatives (FN), and (b) RF(1) with low recall and high precision (2\% of network remaining), showing many false negatives (FN) and relatively few false positives (FP).  Solid lines show predicted paths from source to target.  Dashed lines show particle backbones.}
\end{figure}

In order to evaluate the quality of our classification results, we illustrate two cases on the DFN from Figure~\ref{fig:dfn}.  In Figure~\ref{fig:18p-91r}~(a), we visualize the result of our classifier with the highest recall and lowest precision, SVM(0.90,0.054).  Most of the nodes in the particle backbone are classified as positive. The few false negatives (FN) are near the source, and are primarily fractures intersecting the source plane where high particle concentrations accumulate.  False positives (FP) are far more prevalent, forming many connected source-to-target paths that are not in the particle backbone.  In spite of these, the reduced network identified by the classifier contains only 40\% of the original fractures.

In Figure~\ref{fig:18p-91r}~(b), we visualize the result of our classifier with the highest precision and lowest recall, RF(1).  While we see almost no false positives (FP), most of the nodes in the particle backbone are missed.  The false negatives (FN) near the source are not necessarily of great concern, as these simply represent the inlet plane, but only one connected path exists between source and target.  On some other networks in the test set, the classifier does not even generate a connected source-to-target path at all.  The drastic reduction of network size, to 2\% of the original fractures, results in too much loss of physical relevance.

\begin{figure}[t]
	\centering
	\includegraphics[width=0.8\linewidth]{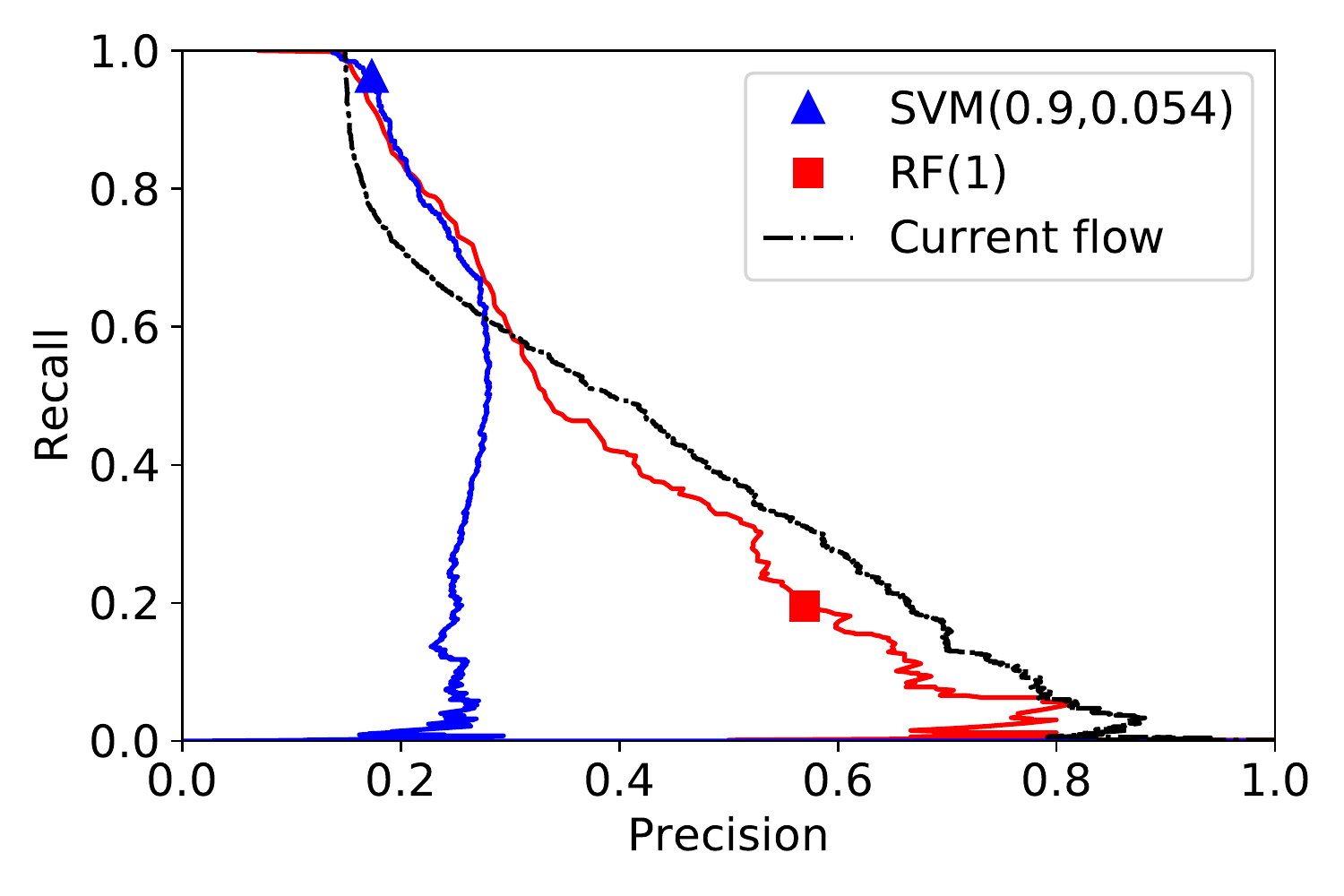}
	\caption{Precision/recall obtainable with one SVM and one RF classifier, along with current-flow thresholding for comparison.  Markers indicate performance of unmodified classifier.}
	\label{fig:prcurve}
\end{figure}

It is instructive to consider the full range of accessible precision and recall values for the classifiers above, as we did for the simple current-flow thresholding method in Section~\ref{S:methods}.  Given a trained classifier with given hyperparameter values $\mathbf{q}$, one can modify it to give more or fewer positive assignments, effectively changing the percentage of votes needed for a positive classification (in the case of RF) or shifting the decision boundary (in the case of SVM).  Note that this is not the same as generating different classifiers from the training data, as in Tables~\ref{table:rf-models} and \ref{table:svm-models}.  Figure \ref{fig:prcurve} shows precision/recall curves generated in this way for the two classifiers used in Figure~\ref{fig:18p-91r}, along with the current-flow curve as a baseline comparison.  Marker values represent the precision/recall values seen in Tables~\ref{table:rf-models} and \ref{table:svm-models} for the unmodified classifiers.  It appears at first that current-flow thresholding has the strongest performance below about 60\% recall.
However, as with RF(1) above, these are nonphysical results: it generates connected subnetworks only for the highest recall values, where it significantly underperforms RF and SVM.

As discussed earlier, our primary objective is not reconstructing training data, but rather reducing network size while maintaining crucial flow properties.  These properties are measured by the breakthrough curve (BTC), which gives the distribution of simulated particles passing through the network from source plane to target plane in a given interval of time.  
This is a common quantity of interest in subsurface systems, where one needs to predict the travel time distribution through the fracture network in order to evaluate the performance of systems such as hydraulic fracturing, nuclear waste disposal or gas migration from a nuclear test. 
We would like the BTC for our reduced networks to match that of the full network in a number of respects, notably the shape of the cumulative distribution function and the fraction of particles that reach the target plane after a given time.

\begin{figure}[!b]
\centering\includegraphics[width=0.8\linewidth]{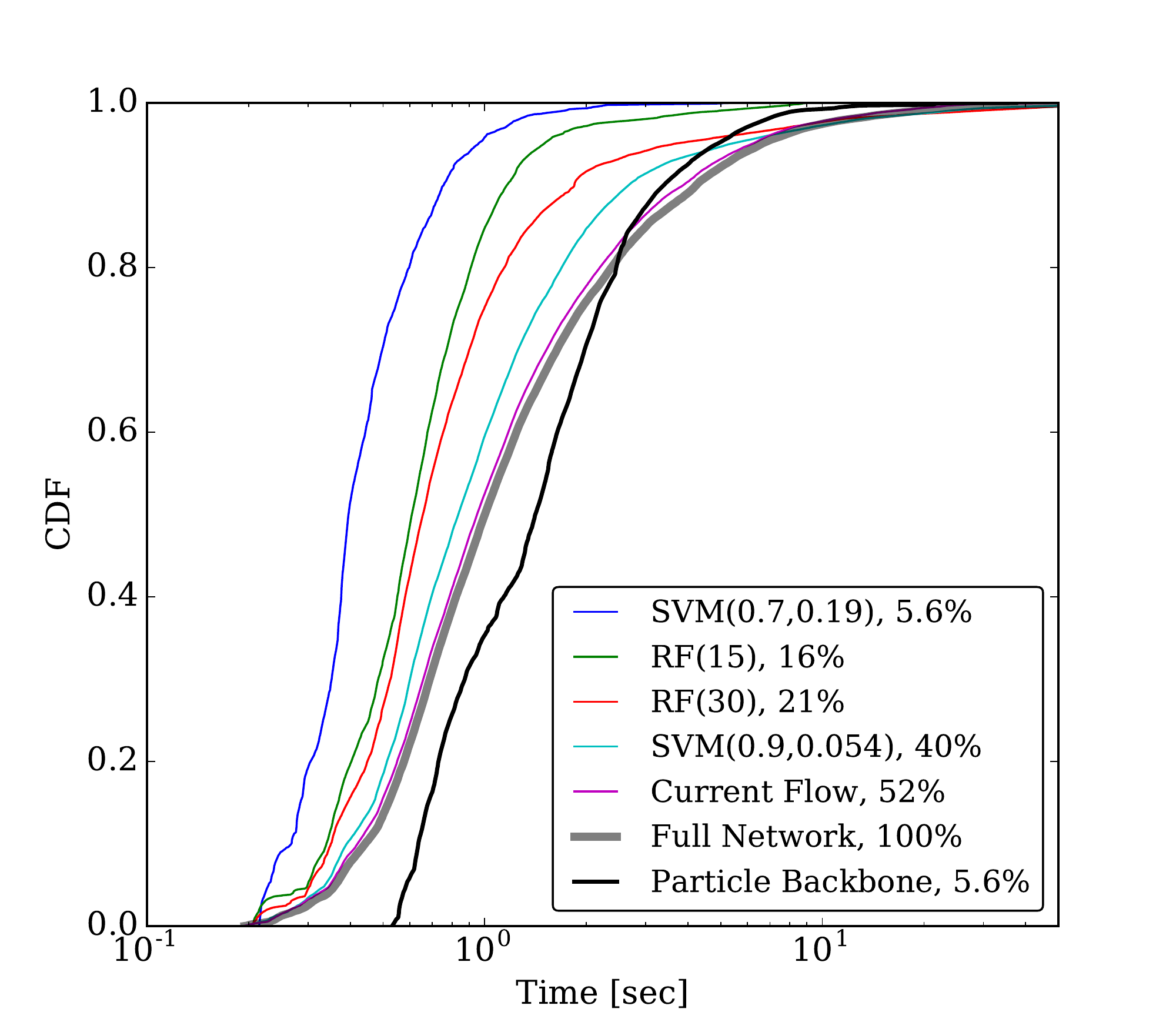}
\caption{Predictions for the DFN from Figure~\ref{fig:dfn}, visualized as BTC (cumulative distribution function) produced by {\sc dfnWorks}.  Representative results from four models are given, together with current-flow thresholding, full network and particle backbone.  Legend shows model parameters and size of reduced network.}
\label{fig:btc}
\end{figure}

Figure~\ref{fig:btc} shows the BTC on this network for a representative sample of four of our classifiers.  As a comparison, we also show the BTC for thresholding on nonzero current flow, as well as for the full network and for the particle backbone.  While current flow thresholding gives a very close match, it only reduces the network to 52\% of its original size.  SVM(0.9,0.054) and RF(30) reduce the network to 40\% and 21\%, while still providing acceptable matches.  The median breakthrough time for RF(30) deviates from that of the full network by approximately the same amount as the particle backbone, though in the opposite direction: it underestimates rather than overestimates the breakthrough time.

Finally, in order to quantify the tradeoff between BTC agreement and network reduction, we calculate the Kolmogorov-Smirnov (KS) statistic, giving a measure of ``distance'' between two probability distributions.  The KS statistic is independent of binning, and most sensitive to discrepancies close to the medians of the distributions, making it particularly suitable for comparing BTCs.  The results are summarized in Table~\ref{table:tabular_compare}.  They confirm that the classifier with highest recall, SVM(0.90,0.054), which reduces the network to 40\% of its original size, has a BTC close to that of the full network (KS statistic 0.10).

\begin{table}[htb]
        \begin{center}
        \begin{tabular}{l|c|c}
            Classifier& Fractures remaining & KS \\
            \hline
            Current flow & 52\% & 0.03 \\
            SVM(0.90,0.054)    & 40\% & 0.10 \\
            RF(1400)    & 38\%  &  0.12 \\
            SVM(0.90,0.063)    & 35\% & 0.12 \\            
            SVM(0.70,0.070)    & 22\% & 0.25 \\
            RF(30)    & 21\% &  0.26 \\
            RF(15)    & 16\% &  0.35 \\
            SVM(0.70,0.190)    & 5.6\% &  0.59 \\
            RF(1)    & 2.0\% &  0.68 \\

        \end{tabular}
        \end{center}
        \caption{Results of applying current-flow thresholding, four RF and four SVM models to the DFN from Figure~\ref{fig:dfn}.  Fractures remaining in network are those identified as positive by classifier.  Values differ slightly from results over entire test set, due to graph-to-graph fluctuations.  KS statistic represents distance between breakthrough curve on reduced network and on full network.}
        \label{table:tabular_compare}
\end{table}

\section{Discussion and Conclusions}
\label{S:conc}


We have presented a novel approach to finding a high-flow subnetwork that does not require resolving flow in the network, and that takes minimal computational time.  The method involves representing a DFN as a graph whose nodes represent fractures, and applying machine learning techniques to rapidly predict which nodes are part of the subnetwork.  We used two supervised learning techniques: random forest and support vector machines.  Once these algorithms have been trained on flow data from particle simulations, they successfully reduce new DFNs while preserving crucial flow properties.  Our algorithms use topological features associated with nodes on the graph, as well as a small number of physical features describing a fracture's properties.  We consider each node as a point in the multi-dimensional feature space, and classify it according to whether or not it belongs to the subnetwork.

By varying at most two parameters of our classifiers, we are able to obtain a wide range of precision and recall values.  These yield subnetworks whose sizes range from 40\% down to 2\% of the original network.  For reductions to as little as 21\% of the original size, the resulting breakthrough curve (BTC) displays good agreement with that of the original network.  We therefore obtain significantly reduced networks that are useful for flow and transport simulations and generated in seconds.  By comparison, the computation time needed to extract the particle backbone is on the order of an hour.  
We use cross-validation to identify the crucial classifier parameters, and show how to use it to tune these parameters, optimizing precision for fixed recall or vice-versa.  To the extent that recall approximates BTC agreement, this can result in maximal network reduction for a given level of accuracy, or maximal accuracy for a given level of network reduction.

In addition to the classification results, the random forest method gives a set of relative importances for the features used.  These importances are determined by permuting the values of a given feature and observing the effect this has on classification performance.  We have found that features based on global topological properties of the underlying graph were significantly more important than those based on geometry or physical properties of the fractures.  This reinforces previous observations that network connectivity is more fundamental to determining where flow occurs in a network than are geometric or hydraulic properties for sparse networks~\cite{hyman2016fracture}.  Quantitatively, the most important of our global topological features is source-to-target current flow, which measures how much of a unit of current injected at the source (representing the inlet plane of the DFN) passes through a given node of the graph.

Indeed, classifying fractures only based on whether they conduct nonzero current flow yields in itself a reasonable graph reduction of around 50\%, with a BTC that very closely matches the original network.  However, this does not generalize to a method allowing arbitrary graph reduction: raising the current-flow threshold above zero reduces the number of fractures, but results in subnetworks that are disconnected and therefore nonphysical.  By contrast, when we use the full set of classification features, we consistently realize a connected subnetwork for all but the lowest recall values.  It is somewhat remarkable that this occurs in spite of our classifiers never explicitly making use of source-to-target paths in the graph.

In principle, the performance of classifiers in this framework depends on the particular geometric and hydrological properties of the fracture network generation parameters and the inferred topological structures. Changing generation parameters will not only result in different geometries, but also different topological properties of the network realizations. Depending on the prescribed distributions of fracture radii, network density, fracture shape, fracture intensity, etc., what features should be considered in the classifier might also change. Thus, it is imperative that the classifiers must be trained using the particular ensemble in which they want to predict the subnetworks.

Finally, some evidence suggests that if one could in fact classify paths rather than nodes, results would improve further.  We are currently exploring a classification method that initially labels fractures at the inlet and outlet planes, and then successively attempts to propagate positive identifications through the network, thereby forming source-to-target paths.  The objective of this method is to generate subnetworks that are far closer to the particle backbone itself.  Thus, the training data would be used not merely to guide the classifier toward useful network reductions, but rather in the more conventional machine learning setting of providing ground truth to be reproduced.  Preliminary tests suggest that such a method may considerably boost precision and recall simultaneously, generating subnetworks whose BTC closely matches the full network but whose size is not much larger than the particle backbone.

\section{Acknowledgments}
The authors are grateful to an anonymous reviewer, for comments and suggestions that helped improve the clarity and readability of the manuscript. This work was supported by the Computational Science Research Center at San Diego State University,
the National Science Foundation
Graduate Research Fellowship Program under Grant No.~1321850, and the U.S. Department of Energy at Los Alamos National Laboratory under Contract No.~DE-AC52-06NA25396 through the Laboratory Directed Research and Development program.
JDH thanks the LANL LDRD Director's Postdoctoral Fellowship  Grant $\#$ 20150763PRD4 for partial support. JDH, HSV, and GS thank LANL LDRD-DR Grant $\#$20170103DR  for support.





\bibliographystyle{spmpsci.bst}
\bibliography{Bibliography.bib}







\section*{Flow Equations and transport simulations}
We assume that the matrix surrounding the fractures is impermeable and there is no interaction between flow within the fractures and the solid matrix. 
Within each fracture, flow is modeled using the Darcy flow. The aperture within each fracture is uniform and isotropic, but they do vary between fractures and are positively correlated to the fracture size~\cite{hyman2016fracture}.
Locally, we adopt the cubic law~\cite{witherspoon1980validity} to relate the permeability of each fracture to its aperture.
We drive flow through the domain by applying a pressure difference of 1MPa across the domain aligned with the x-axis.
No flow boundary conditions are applied along lateral boundaries and gravity is not included in these simulations.
These boundary conditions along with mass conservation  
and Darcy's law are used to form an elliptic partial differential equation for steady-state distribution of pressure within each network
\begin{equation}\label{eq:2Dflow}
\nabla \cdot ( k \nabla P ) = 0~.
\end{equation}
Once the distribution of pressure and volumetric flow rates are determined by numerically integrating (\ref{eq:2Dflow}) the methodology of  Makedonska et al.~\cite{makedonska2015particle} and Painter et al.~\cite{painter2012pathline} and are used to determine the Eulerian velocity field $\vu(\vx)$ at every node in the conforming Delaunay triangulation throughout  each network.

The spreading of a nonreactive conservative solute transported is represented by a cloud of passive tracer particles, i.e., using a Lagrangian approach. 
The imposed pressure gradient is aligned with the $x$-axis and thus the primary direction of flow is in the $x$ direction. 
Particles are released from locations in the inlet plane $\vx_0$ at time $t = 0$ and are followed until the exit the domain at the outlet plane $\vx_L$
The trajectory $\vx(t;\va)$ of a particle starting at $\va$ on $\vx_0$  is given by the advection equation, $\dot{\vx(t; \va)} = \vv(t;\va)$ with  $\vx(0;\va) = \va$ 
where the Lagrangian velocity $\vv(t;\va)$ is given in terms of
the Eulerian velocity $\vu(\va)$ as $
\vv(t;\mathbf \va) = \vu[\vx(t;\va)]$.
The mass represented by each particle $m(\va)$ and the breakthrough time at the outlet plane, $\tau(\vx_L;\va)$ of a particle that has crossed the outlet plane, $\vx_L=(L,y,z)$ is can be combined to compute the total solute mass flux $\psi(t)$ that has broken through at a time $t$, 
\begin{equation}\label{eq:CD}
\Psi(t,\vx_L )  = \frac{1}{M} \int\limits_{\Omega_a} d m(\va) \delta[t - \tau(\vx_L,\va)]~,
\end{equation}
where $\Omega_a$ is the set of all particles. Here mass is distributed uniformly amongst particles, i.e., resident injection is adopted. For more details about the injection mode see Hyman et al.,~\cite{hyman2015influence}.

\section*{Algorithm Description}

\subsection*{Random Forest}

The random forest method is based on constructing a collection of decision trees.
A decision tree \cite{russell2010artificial} is a tree whose interior nodes represent binary tests on a feature and whose leaves represent classifications. An effective way of constructing such a tree from training data is to measure how different tests, also called splits, separate the data.  The \emph{information gain} measure compares the entropy of the \emph{parent} node to the weighted average entropy of the \emph{child} nodes for each split.
The splits with the greatest information gain are executed, and the procedure is repeated recursively for each child node until no more information is gained, or there are no more possible splits.
A limitation of decision trees is that the topology is completely dependent on the training set.  Variations in the training data can produce substantially different trees.

The random forest method~\cite{ho_rf_1995,ho_rf_1998} addresses this problem by constructing a collection of trees using subsamples of the training data.  These subsamples are generated with replacement (bootstrapping), 
so that some data points are sampled more than once and some not at all. The sampled ``in-bag'' data points are used to generate a decision tree. The ``out-of-bag'' observations (the ones not sampled) are then run through the tree to estimate its quality \cite{liaw_classification_2002}. This procedure is repeated to generate a large number (hundreds or thousands) of random trees.  

To classify a test data point, each tree ``votes'' for a result.  This provides not only a predicted classification, determined by majority rule, but also a measure of certainty, determined by the  fraction of votes in favor. The use of bootstrapping effectively augments the data, allowing random forest to perform well using fewer features than other methods. 
The category with more votes is assigned to the new observation. The idea of random decision forests originated with T. Ho in 1995. Ho found that forests of trees partitioned with hyperplanes can have increased accuracy under certain conditions \cite{ho_rf_1995}. In a later work \cite{ho_rf_1998}, Ho determined that other splitting methods, under appropriate constraints, yielded similar results.

Additionally, random forest provides an estimate of how important each individual feature is for the class assignment.  This is calculated by permuting the feature's values, generating new trees, and measuring the ``out-of-bag'' classification errors on the new trees. If the feature is important for classification, these permutations will generate many errors.  If the feature is not important, they will hardly affect the performance of the trees.

\subsection*{Support vector machines}

Support vector machines (SVM) use a \emph{maximal margin classifier} to perform binary classification.  Given training data described by $p$ features, the method identifies boundary limits for each class in the $p$-dimensional feature space.  These boundary limits, which are $(p-1)$-dimensional hyperplanes, are known as local classifiers, and the distance between the local classifiers is called the margin.  SVM attempts to maximize this margin, making the data as separable as possible, and defines the classifier as a hyperplane in the middle that separates the data into two groups.  The data points on the boundaries are called support vectors, since they ``support'' the limits and define the shape of the maximal margin classifier. 

Formally, a support vector machine attempts to construct a hyperplane,
\begin{equation}
\label{eqn:linear_boundary}
\langle \ba, \bx_i \rangle + b = 0~,
\end{equation} 
that partitions data points $\bx_i$ into disjoint sets, for $i\in\{1,\dots,n\}$. 
To define the hyperplane we must determine the unknown coefficients $\ba = (a_1,
\ldots, a_p)$ and $b$. SVM seeks to determine a hyperplane that separates the two sets
$A$ and $B$ and leaves the largest margin.

Let $\langle \ba, \bx_i \rangle + b = \pm 1$ be the normalized equations for the
boundaries of the hyperplane margin. Data points $\bx_i$ on either side of the
margins of the hyperplane will lie in either $A$ and $B$ depending upon
whether they satisfy either
\begin{equation}\label{eq:svm_inequalities}
\langle \ba, \bx_i \rangle  + b > 1\qquad {\text or}  \qquad \langle \ba, \bx_i \rangle  + b < -1~.
\end{equation}
Define an indicator function $S(\bx)$ by
\begin{equation}\label{eq:transitions}
S(\bx) =\begin{cases}
   1 \qquad \bx \in A\\
     -1 \qquad \bx \in B\\
 \end{cases}
\end{equation}
and set $S_i = S(\bx_i)$. Then we can combine (\ref{eq:svm_inequalities}) into
a single inequality $(\langle \ba, \bx_i \rangle + b) S_i \ge 1 $ for all
$\bx_i$; with equality holding for support vectors, the nearest points to the margin.  

In most cases the sets $A$ and $B$ may only be close to linearly separable.  To
account for this possibility we introduce slack variables $\xi_i \ge 0$,
\begin{equation}\label{eq:slack_hyperplane}
(\langle \ba, \bx_i \rangle  + b) S_i \ge 1 - \xi_i \qquad \forall~i~,
\end{equation}
that allows $\bx_i$ corresponding to $\xi_i > 0$ to be incorrectly classified
with the $\xi_i$ being used as a penalty term.  Distances $\rho_1$ and
$\rho_2$ from the coordinate origin to the margin boundary are given by
$\rho_1 = - (b +1)/\| \ba \|$ and $\rho_2 = - (b - 1)/\| \ba \|$ where $\| \ba
\|$ is the Euclidean length of $\ba$.  The margin width is the distance
between these two lines $d = \rho_2 - \rho_1 = {2/\| \ba\|}$.  We seek to
maximize $d$ or, equivalently, minimize $\| \ba \|$, over the set of training data $(\by_1,\dots,\by_m)$, subject to the linear
constraints (\ref{eq:slack_hyperplane}).  Typically one works with the
Lagrangian formulation of the constrained optimization problem.  The dual
Lagrangian problem is to minimize the objective function
\begin{eqnarray}\label{eq:svm_nonseparable}
 \nonumber L(\ba, b, \boldsymbol{ \xi}, \gamma, \boldsymbol{\delta}) &=& \frac{1}{2}{\| \ba\|}^2 - \sum_{j=1}^m \gamma_j [ (\langle \ba, \by_j \rangle + b)  S_j - 1 + \xi_j] \\
 && + C \sum_{j=1}^m \xi_j -  \sum_{j=1}^m \delta_j \xi_j~
\end{eqnarray}
subject to the non-negativity constraints of the Lagrange multipliers
$\gamma_j, \delta_j$, and $\xi_j \ge 0$ and obtain $\ba$ and $b$.  The penalty, or
\emph{margin parameter}, $C$ is a regularization term that controls how many
points are allowed to be mislabeled in the SVM hyperplane construction; smaller values of
$C$ allow for more points to be mislabeled. A
solution to this optimization problem defines $\ba$ and $b$. 
The SVM
classifier is then given by the sign of the \emph{decision function},
\begin{equation}\label{svm_decision_classifier1}
\chi(\bx_i) = \sgn ( \langle \ba, \bx_i \rangle + b).
\end{equation}

%

SVM falls into the category of kernel methods, a theoretically powerful and computationally efficient means of generalizing linear classifiers to nonlinear ones.  For instance, on a two-dimensional surface ($p=2$ features), instead of the line described by equation~(\ref{eqn:linear_boundary}), we can choose a polynomial curve or a radial loop.  Equation (\ref{svm_decision_classifier1}) may then be written in the form
%
\begin{equation}
\chi(\bx_i) = \sgn\left(\alpha_0 + \sum_{j=1}^m\alpha_j K(\bx_i,\by_j)\right),
\end{equation}
where $K(\bx_i,\by_j)$ is an appropriate kernel function of $\bx_i$ and training point $\by_j$.
%
Radial kernels often provide the best classification performance, but at higher computational costs.

\end{document}